\documentclass[aps,pra,twocolumn,superscriptaddress,notitlepage,nofootinbib,longbibliography,10pt]{revtex4-2}
\usepackage[colorlinks=true, allcolors=blue, citecolor=blue]{hyperref}
\usepackage{mathrsfs}
\usepackage{epsfig}
\usepackage{graphicx}
\usepackage{amsfonts}
\usepackage[figuresright]{rotating}
\usepackage{amssymb}
\usepackage{amsmath}
\usepackage{dcolumn}
\usepackage{bm}
\usepackage{float}
\usepackage[dvipsnames]{xcolor}
\usepackage{braket}
\usepackage{units}
\usepackage{xspace}
\usepackage{bbold}
\usepackage{orcidlink}
\usepackage{tabularx}
\usepackage{adjustbox}

\newcommand{\NU}{School of Physics and Technology, Nantong University, Nantong 226019, China}
\newcommand{\ANU}{State Key Laboratory of Opto-Electronic Information Acquisition and Protection Technology, School of Physics, Anhui University, Hefei
		230601,  China}
\newcommand{\JSU}{Department of Physics, Jiangsu University, Zhenjiang 212013, China}
\begin{document}
\title{Finer sub-Planck structures and displacement sensitivity of SU(1,1) circular states}

\author{Naeem Akhtar}
\affiliation{\ANU}
\author{Jia-Xin Peng}
\email{JiaXinPeng@ntu.edu.cn}
\affiliation{\NU}
\author{Tariq Aziz}
\affiliation{\ANU}
\author{Xiaosen Yang}
\affiliation{\JSU}
\author{Dong Wang}
\email{dwang@ahu.edu.cn}
\affiliation{\ANU}
\date{\today}
\begin{abstract}
Quantum states with sub-Planck features exhibit sensitivity to phase-space displacements beyond the standard quantum limit, making them useful for quantum metrology. In the context of the SU(1,1) group, sub-Planck features have been constructed through the superposition of four Perelomov coherent states on the hyperbolic plane (the SU(1,1) compass state). However, these structures differ in scale along different phase-space directions (anisotropic features), resulting in nonuniform sensitivity enhancement to phase-space displacements. Here, we construct $N$-component compass states, which are obtained by superposing $N \geq 6$ SU(1,1) coherent states with an even total number, evenly arranged along a circular path on the hyperbolic plane; that is, all components lie at the same distance from the origin and have equal angular spacing of $\frac{2\pi}{N}$. We observe that these generalized SU(1,1) compass states exhibit isotropic sub-Planck structures, leading to an isotropic enhancement in sensitivity to phase-space displacements that progressively increases with larger $N$. These states are directly relevant to quantum platforms supporting Kerr-type interactions between two bosonic modes, where the underlying SU(1,1) dynamical symmetry enables the generation of multicomponent SU(1,1) compass states. Specifically, the compact evolution of SU(1,1) coherent states under the considered dynamics enables the generation of multicomponent SU(1,1) compass states at specific times. We also investigate the effects of thermal decoherence on these multicomponent compass states within a frequency-resolved Lindblad framework, demonstrating the evolution and degradation of their nonclassical signatures.
\end{abstract}
\maketitle
\section{INTRODUCTION}
The quantum uncertainty principle~\cite{Robertson1929,wheeler2014}, derived from commutator relations such as the position-momentum quadratures $[\hat{x}, \hat{p}] = i\hbar$, imposes a limit on the size of a phase-space structure. In the Wigner phase-space representation associated with the Heisenberg-Weyl (HW) symmetry~\cite{Sch01}, the product of uncertainties in position ($\Delta x$) and momentum ($\Delta p$) estimates the phase space size of a feature, which satisfies the uncertainty relation $\Delta x \Delta p \geq \frac{\hbar}{2}$, implying that a phase space feature cannot be smaller than the Planck scale ($\hbar$). One might question the possibility of phase-space features smaller than the Planck scale (sub-Planck structures). While such sub-Planck features challenge the uncertainty principle and raise doubts about their physical implications, these structures appear to have tangible physical consequences. In simpler terms, they may exist~\cite{Zurek2001,Praxmeyer007,Howard2019,PhysRevLett.89.154103}, and are indeed physical~\cite{zurek_RevModPhys2003}, with their role being particularly crucial in quantum metrology~\cite{Toscano06,Eff4,zheng2025,deng2024quantum,PhysRevA.93.053835}. Decoherence can pose a significant challenge in preserving the quantum signatures of states~\cite{PhysRevLett.113.140401,PhysRevLett.134.120201}. Specifically, quantum states with finer-scale phase-space features appeared highly fragile under decoherence~\cite{akhtar2025D,zurek_RevModPhys2003,Akhtar2024}.

Sub-Planck structures have been created from diverse quantum states, particularly within the HW group~\cite{PhysRevA.109.033724,arman2026enhancingsizephasespacestates,naeemprapplied,llhr-hn2y,naeem2023,moore2026}, with less attention given to analogs in other symmetry groups~\cite{Naeem2021,naeem2022}. The SU(1,1) group has shown considerable utility in quantum metrology~\cite{Du2020, rv9m-n7px, Hudelist2014, Berrada2013, Szigeti2017, Barry2020, Linnemann_2017, h2x6-dz96}, nonclassical state generation~\cite{morrison2023, Klimov2021, Seyfarth2020}, and two-photon effects~\cite{Gerry1991, Gerry1995}, to name a few. The superposition of four Perelomov SU(1,1) coherent states on a hyperboloid surface, referred to as the SU(1,1) compass state~\cite{naeem2022}, can also be represented on the stereographic plane (Poincar\'e\xspace disk), which gives rise to sub-Planck features of the SU(1,1) group. 

The enhancement in sensitivity to phase-space displacements beyond the standard quantum limit (SQL) offered by sub-Planck features depends on their structural form, with different sensitivities held by their anisotropic and isotropic forms~\cite{akhtar2024sub-shot,naeemprapplied,llhr-hn2y,Barry2023}. Anisotropic sub-Planck structures yield direction-dependent enhancements in sensitivity to phase-space displacements, whereas isotropic counterparts exhibit a higher degree of sub-Planckness and enable phase-space sensitivity with greater precision~\cite{naeemprapplied}. These properties make isotropic sub-Planck features particularly attractive for quantum metrology. To the best of our knowledge, isotropic sub-Planck features within the SU(1,1) group have not yet been realized. In this work, we aim to develop refined sub-Planck features in the SU(1,1) framework by constructing multicomponent SU(1,1) compass states. Specifically, these generalized versions of the SU(1,1) compass state are constructed from the superposition of $N$ Perelomov SU(1,1) coherent states on the Poincar\'e\xspace disk, where $N$ ranges from small values ($N = 6$) to large values ($N \to \infty$). In these superpositions, the coherent states are symmetrically arranged along a circular path on the Poincar\'e disk, with the total number $N$ restricted to even values; each component state lies at the same distance from the origin, and adjacent states are separated by an angle of $\frac{2\pi}{N}$. The resulting multicomponent SU(1,1) compass states generate isotropic sub-Planck structures, and as the number of superposed coherent states increases, the isotropy of these features becomes more pronounced. Furthermore, our generalized SU(1,1) compass states exhibit isotropic enhancement in their sensitivity to phase-space displacements, resulting in precision that exceeds the SQL consistently across all displacement directions, making them valuable resources for quantum metrology applications.

The Hamiltonian supports different symmetry groups that may usually appear in interacting multi-mode systems~\cite{PhysRevD.29.1107,Lawrie2029,Bu1990,SOLOMON}. It has been observed that the Hamiltonian, corresponding to Kerr-type interactions between two bosonic modes and exhibiting SU(1,1) symmetry, can evolve initial SU(1,1) coherent states under temporal compact evolution into superpositions of SU(1,1) coherent states~\cite{quantum7030031}. We note that this system can generate higher-component SU(1,1) superpositions, leading to the the multicomponent SU(1,1) compass states. Finally, we discuss the stability of our multicomponent SU(1,1) compass states under decoherence modeled by a frequency-resolved thermal Lindblad channel. Our results indicate that SU(1,1) compass states with a higher number of components are comparatively more delicate under decoherence due to their more fragile phase-space features.

This paper is organized as follows: Sec.~{\ref{sec:main_conc}} reviews the concept of sub-Planck structures and their connection to displacement sensitivity associated with the quantum states of the HW group. Sec.~\ref{sec:su(1,1)_basics} reviews basics of the SU(1,1) group. Sec.~\ref{sec:notable} introduces multicomponent SU(1,1) compass states exhibiting isotropic sub-Planck structures, discusses their generation, investigates their displacement sensitivity, and analyzes their stability under decoherence. Finally, Sec.~\ref{sec:summary} summarizes the main results of this work.

\section{MAIN THEORY}\label{sec:main_conc}

To make the discussion as self-contained as possible, we begin with a brief review of the main concepts underlying the present investigation, specifically focusing on the sub-Planck structures and their impact in quantum measurements. Nonclassical states play a crucial role in quantum metrology~\cite{Sinatra2022,Czartowski_2024,Martin_2020,RevModPhys.90.035005,PhysRevLett.128.150501,7pyw-tgjd}, and states exhibiting sub-Planck structures are evident as valuable resources for quantum metrology, both theoretically~\cite{Toscano06} and experimentally~\cite{zheng2025}. 

In more specific cases, the usefulness of a quantum state to detect phase space displacement or rotations (i.e., phase shifts) can be quantified by analyzing how distinguishable it becomes from its original form after a minor translation~\cite{Toscano06}. This reflects the sensitivity to phase-space displacements and is closely related to the phase-space scales, where quantum states with sub-Planck-scale structures are particularly desirable~\cite{Naeem2021,naeem2022,naeem2023,naeemprapplied,akhtar2024sub-shot}. 
To understand this, let us start with quantum states belonging to the well-established HW group, which is based on $\mathfrak{hm}$(1) algebra, defined through the annihilation operator $\hat{a}$ and the creation operator $\hat{a}^\dagger$, which satisfy the canonical commutation relation $[\hat{a}, \hat{a}^\dagger] = 1$, with $\hbar = 1$ hereafter. 

The Wigner function for a generic quantum state $\hat{\rho}$ belonging to the HW group can be expressed as
\begin{align}\label{eq:hw_wigner}
W_{\hat{\rho}}(x,p)=\mathrm{tr}[\hat{\rho}\hat{D}(x,p)\hat{\Pi}\hat{D}^\dagger(x,p)],
\end{align}
where $\hat{D}(x,p)=\hat{D}\left(\frac{x+\text{i}p}{\sqrt{2}}\right)$ with $\hat{D}(\alpha)=\exp\left(\alpha\hat{a}^\dagger-\alpha^*\hat{a}\right)$ being the displacement operator, $\hat{\Pi}=(-1)^{\hat{a}^\dagger\hat{a}}$ denoting the parity operator, with $x$ and $p$ representing the canonical position–momentum pair. When parity measurements are available, Eq.~(\ref{eq:hw_wigner}) gives a mechanism for experimentally measuring the Wigner function, as demonstrated in systems such as cavity QED and trapped ions~\cite{PhysRevLett.78.2547}. However, in some cases, parity measurements can be difficult, necessitating the use of other observables. For example, in standard measurement protocols for propagating optical fields, homodyne detection is typically utilized, and the Wigner function can be reconstructed using tomographic techniques~\cite{RevModPhys.81.299}.

The overlap between a state and its slightly displaced version quantifies the phase space sensitivity of a state~\cite{Audenaert14,Toscano06}. For a generic state $\hat{\rho}$ and its $\delta$-perturbed version, the overlap can be expressed as $S_{\hat{\rho}}(\delta)=\text{tr}[\hat{\rho}\hat{D}(\delta)\hat{D}^\dagger(\delta)]$, which for any pure state $\ket{\psi}$ takes the form $S_{\ket{\psi}}(\delta)=|\braket{\psi\hat{D}(\delta)\psi}|^2$, where $\ket{\psi^\prime}=\hat{D}(\delta)\ket{\psi}$ represents slightly displaced counterparts, with $\delta=\delta x+\text{i}\delta p$ being the phase-space displacements. Alternatively, this overlap can be expressed in terms of corresponding Wigner functions~\cite{PhysRevA.69.052111}, which relate the sensitivity with phase-space scales via
\begin{align}\label{eq:wig_sens}
    S_{\ket{\psi}}(\delta)=\frac{2}{\pi}\int^{\infty}_{-\infty} dx dp W_{\ket{\psi}} W_{\ket{\psi^\prime}}.
\end{align}
The phase space area of a coherent state $\ket{\alpha}$ adheres to the minimal uncertainty bound, with the overlap function given by $S_{\ket{\alpha}}(\delta) = \mathrm{e}^{-\left|\delta\right|^2}$, for which distinguishability is achieved when $|\delta| > 1$, indicating that the coherent state is responsive to displacements at the SQL (shot noise limit), thereby setting its resolution as a metrological tool~\cite{Eff4}. This led to the development of quantum states with finer phase-space features compared to coherent states, which may exhibit sensitivity greater than the SQL, often known as sub-shot noise sensitivity~\cite{Toscano06}. 

The type of perturbation in Eq.~(\ref{eq:wig_sens}) results in a phase-space translation ($\delta$) of the states. This form of perturbation has been commonly employed to assess the sensitivity of quantum states~\cite{naeem2022}. Its use in this work allows for a direct comparison of the metrological potential of the present states with that of their previous counterparts. Other types of perturbations, such as rotation angles or the phase of a displacement, can also be considered~\cite{Fadel_2025}. From an experimental perspective, quantum metrology was initially implemented primarily in optical setups. Modern techniques such as microwave photons~\cite{vlastakis2013deterministically}, motional states of trapped ions~\cite{leibfried2005creation}, and solid-state oscillators~\cite{ockeloen2018stabilized} now ease the implementation of complex quantum systems for quantum metrology.

Let us start with the example of the HW cat state, which is the superposition of two distinct coherent states \big($\ket{\text{cat}}\sim\Ket{\frac{x_0}{\sqrt{2}}}+\Ket{-\frac{x_0}{\sqrt{2}}}$ with $x_0 \in \mathbb{R}$\big), may develop interference fringes with the same scale as the coherent state along the position ($x$) direction in phase space, but along the momentum direction ($p$), their extension is proportional to $\frac{1}{x_0}$, leading to directional fineness in phase space. This scenario directly corresponds to phase space sensitivity: a cat state with anisotropically finer phase-space features may also display an anisotropic enhancement in sensitivity compared to a coherent state~\cite{naeemprapplied}. 

The HW compass state \Big($\ket{\text{compass}}\sim\Ket{\frac{x_0}{\sqrt{2}}}+\Ket{-\frac{x_0}{\sqrt{2}}}+\Ket{\frac{\text{i}x_0}{\sqrt{2}}}+\Ket{-\frac{\text{i}x_0}{\sqrt{2}}}$\Big) exhibits sub-Planck features~\cite{Zurek2001}, which are phase-space features that are simultaneously localized along $x$ and $p$ directions, with a nearly uniform scaling of $\frac{1}{x_0}$~\cite{Naeem2021}. This results in sensitivity enhancement dedicated to all phase-space directions, eventually providing greater sensitivity compared to the cat state or coherent state. This enhanced sensitivity of the compass state regards it as a potential choice in quantum metrology~\cite{Eff4,Toscano06}, and it has also been made accessible in real experimental setups~\cite{zheng2025}. 

Subsequently, quantum states with finer sub-Planck features can attain greater sensitivity than compass states~\cite{akhtar2024sub-shot,naeemprapplied}. Specifically, multicomponent HW compass states can be produced by superimposing a large number of coherent states in a circular pattern in phase space, resulting in isotropic sub-Planck structures, which hold more metrological potential compared to their anisotropic counterparts found in the traditional HW compass state~\cite{Zurek2001}. 
The notion of coherent-state superpositions has been expanded to other symmetries~\cite{Per86, Berrada2013, ma2011quantum, Yazdi2008}, and compass states with sub-Planck features have been generated beyond their conventional HW counterparts~\cite{Naeem2021, naeem2022}.

\section{GENERAL SETUP OF THE SU(1,1) GROUP} \label{sec:su(1,1)_basics}

The irreducible representations of SU(1,1) are defined by the eigenvalues of the Casimir operator~\cite{Per86}, namely
\begin{equation}
\hat{K}^2 = \hat{K}_0 - \frac{1}{2}(\hat{K}_{+}\hat{K}_{-} + \hat{K}_{-}\hat{K}_{+}) = k(k-1)\mathbb{1}   
\end{equation}
with $\hat{K}_+$, $\hat{K}_-$, and $\hat{K}_0$ being the generators of the $\mathfrak{su}(1,1)$ Lie algebra and satisfying the commutation relations $[\hat{K}_0, \hat{K}_{\pm}] = \pm \hat{K}_{\pm}$. Furthermore, the generators $\hat{K}_{\pm}$ can be expressed in terms of the Hermitian operators $\hat{K}_1$ and $\hat{K}_2$ as $\hat{K}_{\pm} = \pm \text{i} (\hat{K}_1 \pm \text{i} \hat{K}_2)$. 
The action of the $\mathfrak{su}(1,1)$ generators on the Fock space states $\{\ket{k,n}; n \in \mathbb{N}\}$ is
\begin{align}
&\hat{K}_0\ket{k,n}=(k+n)\ket{k,n}, \\&\hat{K}_{+}\ket{k,n}=\sqrt{(n+1)(2k+n)}\ket{k,n+1}, \\&\hat{K}_{-}\ket{k,n}=\sqrt{n(2k+n+1)}\ket{k,n-1},
\end{align}
where $k$ is the Bargmann index, which is restricted to the positive discrete series. 

The SU(1,1) displacement operator is expressed as~\cite{Per86}:

\begin{align}
\hat{D}(\zeta)=&\nonumber\text{e}^{\xi \hat{K}_+-\xi^* \hat{K}_-},\\=&\text{e}^{\zeta \hat{K}_+}\text{e}^{\ln(1-\left|\zeta\right|^2) \hat{K}_0}\text{e}^{-\zeta^* \hat{K}_-},
\end{align}
where $\xi = \frac{\tau}{2} \text{e}^{\text{i}\varphi}$ represents the points on the upper sheet of the two-sheet hyperboloid surface, and $\zeta = \text{e}^{\text{i}\varphi} \tanh\left(\frac{\tau}{2}\right)$, where $-\infty < \tau < \infty$ and $0 < \varphi < 2\pi$, which alternatively represents the stereographic projection of the upper sheet of the two-sheet hyperboloid onto the unit Poincar\'e\xspace disk with $\zeta$ confined to $0 \leq |\zeta| < 1$.

The Premelov SU(1,1) coherent states are obtained by applying the displacement operator $\hat{D}(\zeta_0)$ to the reference state $\ket{k,0}$~\cite{Per86, Gerry1991}, that is,
\begin{align}\label{eq:su(1,1)_coh}
\ket{\zeta_0,k} = &\nonumber\hat{D}(\zeta_0) \ket{k,0},\\=&\left(1-|\zeta_0|^2\right)^k \sum_{n=0}^{\infty}\left(\frac{\Gamma(n+2 k)}{n!\Gamma(2 k)}\right)^{1 / 2} \zeta_0^n\ket{k, k+n}.
\end{align}
Alternatively, the SU(1,1) coherent states can be associated with points on the two-sheeted hyperboloid surface, given by $\bm{n} = (\cosh\tau, \sinh\tau\cos\varphi, \sinh\tau\sin\varphi)$, representing the hyperbolic counterpart of the Bloch vector. The association between the SU(1,1) groups and the bosonic system belonging to the HW group can be established by the single- and two-mode bosonic realization of the SU(1,1)~\cite{Gerry01}.

\begin{figure}[htp!]
\centering
\includegraphics[width=1.015\columnwidth]{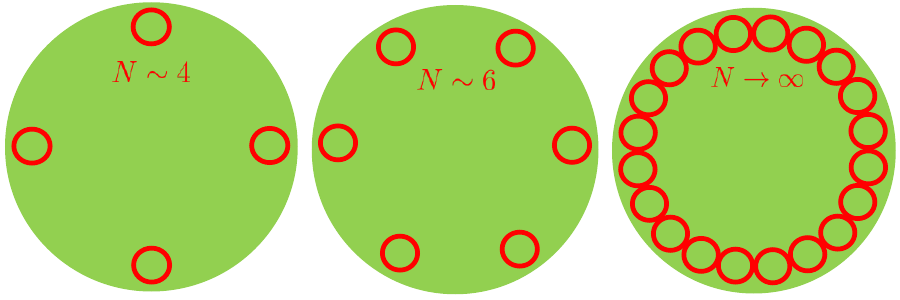}
\caption{SU(1,1) circular states are the superposition of $N$ SU(1,1) coherent states (red circles), which are evenly distributed along a circular path in the Poincar\'e disk (the SU(1,1) phase space). We focus on superpositions with higher $N$, restricted to even integers, where each adjacent component is separated by an equal angular spacing of $\frac{2\pi}{N}$. In the limit $N \to \infty$, the superposition approaches a continuous circular form.}
\label{fig:fig1}
\end{figure}

In the case of single-mode bosonic realization, corresponding generators can be expressed in terms of $\hat{a}^\dagger$ and $\hat{a}$ operators as follows:
\begin{equation}
\hat{K}_+=\frac12\hat{a}^{\dagger2},\, \hat{K}_-=\frac12\hat{a}^2,\,\hat{K}_0=\frac14\left(\hat{a}\hat{a}^{\dagger}+\hat{a}^{\dagger}\hat{a}\right).
\end{equation}
The Casimir operator gets the form $\hat{K}^2 = -\frac{3}{16}$, which associates two Bargmann index ($k$) values~\cite{Gerry01}. For $k = \frac{1}{4}$, the SU(1,1) coherent state leads to the typical squeezed-vacuum state of the form given by 
\begin{equation}
\Ket{\zeta_0, \frac{1}{4}} = (1 - |\zeta_0|^2)^{\frac{1}{4}} \sum_{n=0}^{\infty}\frac{(2n!)^{\frac{1}{2}} \zeta_0^n}{2^n n!}\ket{2n},
\end{equation}
and $k = \frac{3}{4}$ corresponds to 
\begin{equation}
\Ket{\zeta_0, \frac{3}{4}} = (1 - |\zeta_0|^2)^{\frac{3}{4}} \sum_{n=0}^{\infty} \frac{[(2n+1)!]^{\frac{1}{2}} \zeta_0^n}{2^n n!} \ket{2n+1},
\end{equation}
which represents the one-photon squeezed state~\cite{Gerry01, PhysRevA.51.1706}.

Let $\hat{a}_1$ ($\hat{a}_1^\dagger$) and $\hat{a}_2$ ($\hat{a}_2^\dagger$) be the annihilation (creation) operators for modes 1 and 2, respectively, and $\ket{n_1}$ and $\ket{n_2}$ denote the number state of these modes. The complete number-state basis for the two-mode field is denoted as $
\ket{n_1,n_2}=\ket{n_1}\otimes \ket{n_2}$, and the corresponding generators get the following form: 
\begin{equation}
\hat{K}_+=\hat{a}_1^\dagger \hat{a}_2^\dagger,~\hat{K}_-=\hat{a}_1 \hat{a}_2,~\hat{K}_0=\frac12(\hat{a}_1^\dagger \hat{a}_1+\hat{a}_2^\dagger \hat{a}_2+1).
\end{equation}
Their action on $\ket{n_1,n_2}$ satisfies 
\begin{align}
&\hat{K}_0\ket{n_1,n_2}=\frac12\left(n_1+n_2+1\right)\ket{n_1,n_2}, \\ &\hat{K}_+\ket{n_1,n_2}=\sqrt{(n_1+1)(n_2+1)}\ket{n_1+1,n_2+1}, \\&\hat{K}_-\ket{n_1,n_2}=\sqrt{n_1 n_2}\ket{n_1-1,n_2-1}.
\end{align}
The Casimir operator $\hat{K}^2_0$ obeys $\hat{K}^2_0=\frac14\left(\eta^2-1\right)$, where $\eta = \hat{a}_1^\dagger \hat{a}_1 - \hat{a}_2^\dagger \hat{a}_2$, and its eigenvalue corresponds to the difference in the number of quanta between modes 1 and 2, i.e., $n_1 - n_2$. The basis vector~$\ket{k,n}$ takes the form
\begin{align}\label{eq:irr}
k=\frac12(\Delta+1),\, n=\frac12\left(n_1+n_2-\Delta\right),
\end{align}
where~$\Delta$ is the degeneracy parameter, representing the eigenvalue of $|\eta|$, which quantifies the asymmetry in the photon numbers between the two coupled modes. Consider a specific scenario when $n_1 = n_2 + \Delta$ and $n_2 = n = 0, 1, 2, \dots$, implying that mode 1 has $\Delta$ more photons than mode 2. The corresponding SU(1,1) Perelomov coherent states can be expressed in terms of the two-mode squeezed number states,
\begin{align}
    \ket{\zeta_0,\Delta}=(1-|\zeta_0|^2)^{\nicefrac{1+\Delta}{2}}\sum^\infty_{n=0}\sqrt{\frac{(n+\Delta)!}{n! \Delta!}}\zeta_0^n \ket{n,n+\Delta},
    \label{eq:coherent_state_twophotons}
\end{align}
which for $\Delta=0$ is the two-mode squeezed vacuum state~\cite{Gerry01}.

\begin{figure*}[htp!]
\centering
\includegraphics[width=\textwidth]{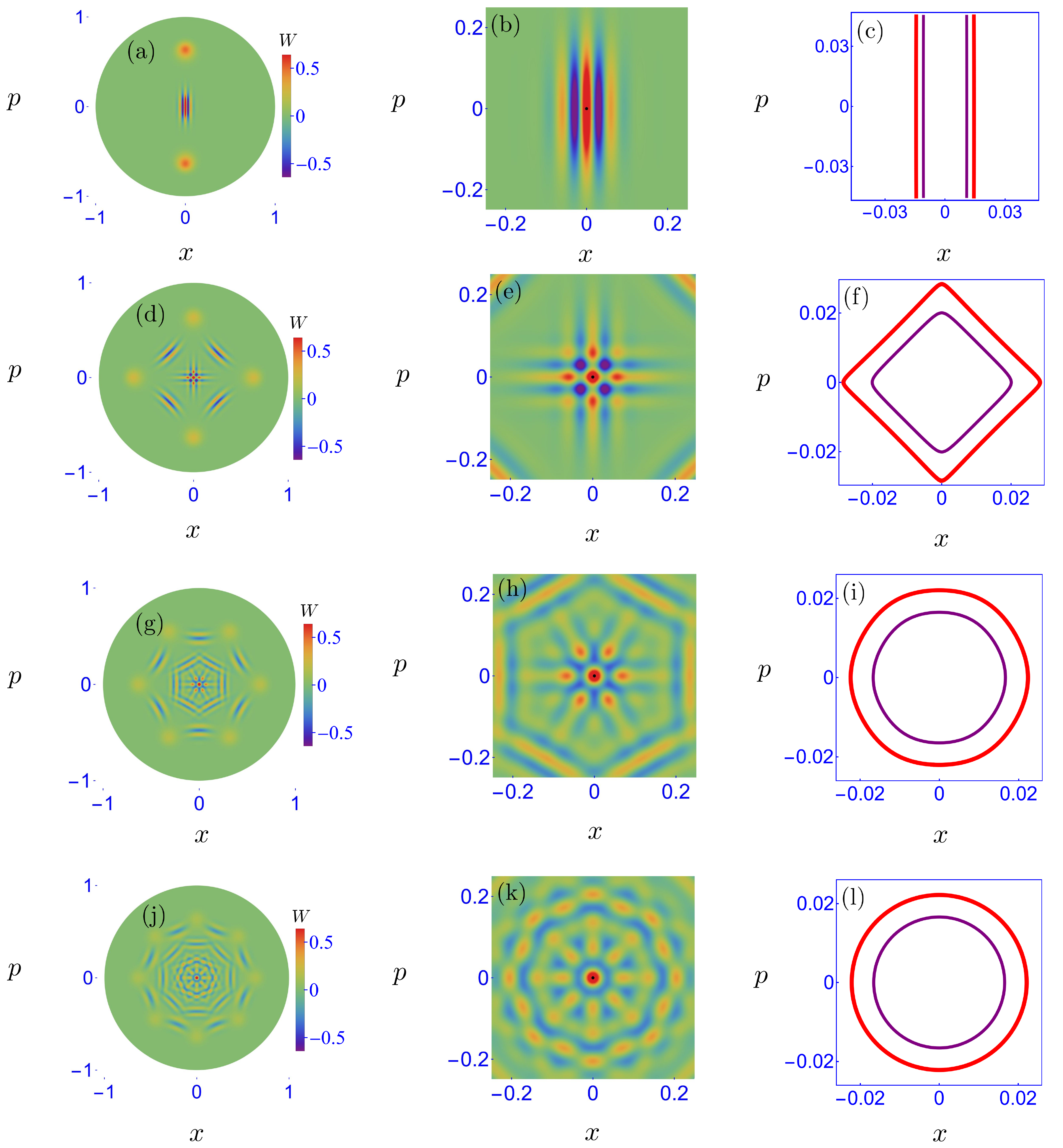}
\caption{The Wigner distributions of the SU(1,1) compass states are shown on the Poincar\'e\xspace plane, with the central focused feature marked by a black dot. (a)-(c) depict the specific instance of the SU(1,1) cat state: (a) the Wigner distribution using $k=12$, (b) highlights the central feature, and (c) displays the zeros associated with the central feature, with red and purple curves corresponding to $k=12$ and $k=16$, respectively. (d)-(f) represent the cases associating superposition of $N=4$ SU(1,1) coherent states, which represents the SU(1,1) compass state, whose Wigner distribution is shown in (d). The central closeup of the Wigner phase space is illustrated in (e) for $k=12$, and the zeros associated with this feature in (f), again with red and purple curves corresponding to $k=12$ and $k=16$, respectively. The case $N=6$ is represented in (g)–(i), where (g) is the Wigner distribution for $k=12$, (h) emphasizes the central feature, and (i) depicts the zeros corresponding to this feature (red curve for $k=12$, purple curve for $k=16$). Finally, for $N=8$ (j)–(l), (j) the Wigner distribution employing $k=12$, (k) the central feature, and (l) the zeros associated with it for $k=12$ (red curve) and $k=16$ (purple curve). In all cases, $\overline{\tau}=1.5$.}
\label{fig:fig2}
\end{figure*}

\section{SU(1,1) SUPERPOSITIONS}\label{sec:notable}

The superposition of SU(1,1) coherent states can give rise to intriguing quantum states~\cite{Klimov2021, naeem2022}; however, surprisingly, such superpositions and their phase space analysis have received little attention, particularly for cases involving a larger number of coherent states, which remain largely unexplored. Similar to the HW group~\cite{Sch01}, the SU(1,1) Wigner function can be expressed as the expectation value of the displaced parity operator~\cite{Seyfarth2020, Klimov2021}, denoted by
\begin{equation}
W_{\hat{\rho}}\left(\zeta\right)=\text{tr}[\hat{\rho}\hat{\Pi}(\zeta)], 
\end{equation}
where 
\begin{equation}
\hat{\Pi}(\zeta)=\hat{D}(\zeta) \hat{\Pi}\hat{D}^{\dagger}(\zeta) 
\end{equation}
with $\hat{\Pi}=\text{e}^{\text{i}\pi(\hat{K}_0-k)}$ is the displaced parity operator. For a generic SU(1,1) operator $\hat{\rho} = \ket{\zeta_i}\bra{\zeta_j}$, this leads to
\begin{align}\label{eq:su11_generic_wig}
W_{\ket{\zeta_i}\bra{\zeta_j}}(\zeta)=\text{exp}\bigg[{2\text{i}k\text{arg}\left(\frac{A}{B}\right)}\bigg]\Bra{\frac{\zeta_j-\zeta}{A}}\hat{\Pi}\Ket{\frac{\zeta_i-\zeta}{B}},
\end{align}
where $A=1-\zeta_j\zeta^*$ and $B=1-\zeta_i\zeta^*$, which in the case of the SU(1,1) coherent state $\ket{k,0}$ gets the form
\begin{equation}
W_{\ket{k,0}}(\zeta) = \sqrt{\Bigg(\frac{|\zeta|^2 - 1}{|\zeta|^2 + 1}\Bigg)^{4k}} \simeq \text{e}^{-4k|\zeta|^2}~\text{for}~k \gg 1.
\end{equation}
This suggests that the spatial extension of the SU(1,1) coherent state along any direction in phase space is proportional to $\frac{1}{\sqrt{k}}$.

Isotropic sub-Planck features of the HW group have been realized through establishing multicomponent compass states~\cite{naeemprapplied}. Here, we construct analogous isotropic sub-Planck structures using multicomponent SU(1,1) compass states. Specifically, we consider superpositions of SU(1,1) coherent states, which are arranged in a specific configuration on the Poincar\'e\xspace disk, as illustrated in Fig.~\ref{fig:fig1}. These superposed components form a circular pattern on the Poincar\'e\xspace disk for higher values of $N$, which can be mathematically expressed as follows:
\begin{equation}\label{eq:su11_generic}
    \ket{\bigcirc_{N}} = \sum_{j=0}^{N-1} \Ket{\text{e}^{\nicefrac{2\pi \text{i} j}{N}}\text{tanh}\left(\frac{\overline{\tau}}{2}\right)}.
\end{equation}
For simplicity, the normalization factors of the states are omitted in this section. The Wigner function of this state can be obtained by using Eq.~(\ref{eq:su11_generic_wig}), which is mathematically denoted as follows:
\begin{align}\label{eq:generic_wig}
    W_{\ket{\bigcirc_{N}}}(\zeta) = \sum_{i,j=0}^{N-1} W_{\ket{\zeta_i}\bra{\zeta_j}}(\zeta), \quad \text{where} \quad \zeta = x + \text{i}p.
\end{align}
Here, $x$ and $p$ represent specific directions on the Poincar\'e\xspace disk. Different values of $N$ correspond to different superpositions, each with distinct phase-space features. The terms in Eq.~(\ref{eq:generic_wig}) with $i=j$ constitute the Wigner function of the composite coherent states (diagonal entries in the density matrix), while the off-diagonal terms ($i\neq j$) of the density matrix correspond to the interference fringes in the phase space. In the following, we examine the phase-space structures of the superpositions considered in this work via their Wigner distributions, with each distribution normalized to its corresponding maximum value.

\subsection{Anisotropic features}

Let us start with the simplest example of the SU(1,1) cat state~\cite{Klimov2021, naeem2022}, aligned along the $p$ axis of the Poincar\'e disk, whose Wigner function is illustrated in Figs.~\ref{fig:fig2}(a)-\ref{fig:fig2}(c). Fig.~\ref{fig:fig2}(a) represents the corresponding Wigner distribution, with interference fringes appearing around the phase space origin, which are further emphasized in Fig.~\ref{fig:fig2}(b). Fig.~\ref{fig:fig2}(c) represents the boundary of the central patch (marked by a black dot in Fig.~\ref{fig:fig2}(b)), specifically obtained by approximating zeros of the central patch under variable values of $k$. Here, lower (red) to higher $k$ (purple) indicates an inverse relationship between $k$ and the extension of the feature~\cite{naeem2022}. This depiction also clearly outlines the edges of the feature. It has been noted that this highlighted central feature has the same extension along $p$ as those of the coherent state (extension along $p$ is proportional to $\frac{1}{\sqrt{k}}$), but is more shrunk by a factor of $\frac{1}{\sqrt{k}}$ along $x$ than that of the coherent state, suggesting greater fineness along a single direction in phase space when compared to the coherent state~\cite{naeem2022}.

Let us now discuss other higher-component states. When $N = 4$ in Eq.~(\ref{eq:su11_generic}), it results in the SU(1,1) compass state~\cite{naeem2022}, consisting of four coherent states, which are equidistant from the origin and uniformly placed on the Poincar\'e\xspace disk. The corresponding Wigner function of this compass state can be obtained from Eq.~(\ref{eq:generic_wig}), which is illustrated in Fig.~\ref{fig:fig2}(d). Apparently this Wigner distribution adopts a tilted square-like shape on the Poincar\'e\xspace disk, with a variety of phase-space features contained inside it, such as the yellow lobes at the corners representing the component coherent states. Between each component state, catlike interference fringes emerge, creating a striking billboard-like pattern around the origin of the Poincar\'e\xspace disk. This central pattern is further shown in Fig.~\ref{fig:fig2}(e), where a single tile, marked with a black dot, is selected for further illustration. 

In Fig.~\ref{fig:fig2}(f), we plot the zero-amplitude points of the central tile for different values of $k$, highlighting the regions where it is confined. This central tile reflects the main sub-Planck feature of the SU(1,1) compass state. Its phase-space extension is roughly proportional to $\frac{1}{k}$ in any direction, exhibiting a greater fineness than those of cat states, but as emphasized in Fig.~\ref{fig:fig2}(f), this resultant central sub-Planck feature still exhibits notable anisotropy, and this anisotropy continues as the state parameters change, especially with different $k$ values, as shown by the shift from red to purple curves in Fig.~\ref{fig:fig2}(f). Such an anisotropic nature of sub-Planck features may put limits on their application in quantum measurements~\cite{Toscano06}.

\subsection{Isotropic sub-Planck features}\label{subsec:isotropic_ideal}

We now examine superpositions involving a larger number of coherent states $N$ in Eq.~(\ref{eq:su11_generic}), focusing on the cases $N=6$ and $N=8$. Setting $N = 6$ in Eq.~(\ref{eq:su11_generic}) results in a six-component compass state, and its associated Wigner function can be obtained by setting $N = 6$ in Eq.~(\ref{eq:generic_wig}), which is depicted on the Poincar\'e\xspace disk, as shown in Fig.~\ref{fig:fig2}(g), where component coherent states are clearly apparent (yellowish lobes), with catlike interference fringes and a kaleidoscopic-like pattern appearing around the origin, where outer features of this distribution give the impression of a hexagon on the Poincar\'e\xspace disk. In Fig.~\ref{fig:fig2}(h), the central portion of the Poincar\'e\xspace disk is emphasized, where the feature of interest is highlighted with a black dot. Fig.~\ref{fig:fig2}(i) highlights the boundary of the central feature, which is obtained by approximating its associated zero-amplitude regions, displaying a nearly circular shape that becomes more localized as $k$ increases (as demonstrated in the red-to-purple curves of Fig.~\ref{fig:fig2}(i)). This evidences the shape of the central feature being nearly isotropic in phase space, with its extension being inversely proportional to $k$ in all directions of phase space.

When comparing the central sub-Planck feature of the six-component compass state illustrated in Fig.~\ref{fig:fig2}(i) with that of the compass state shown in Fig.~\ref{fig:fig2}(f), greater isotropy is associated with the feature of the six-component compass state, indicating a notable improvement in sub-Planckness towards higher-component compass states. Note that the isotropic, localized features that simultaneously localize in all directions of phase space are more in line with the definition of the sub-Planck feature. This can be succinctly described as a higher level of sub-Planckness being attained compared to the compass state.

Setting $N = 8$ in Eq.~(\ref{eq:su11_generic}) leads to a superposition of eight coherent states, establishing our eight-component compass state. The Wigner function of this compass state can be derived by substituting $N = 8$ into Eq.~(\ref{eq:generic_wig}). We illustrate the corresponding Wigner distribution in Fig.~\ref{fig:fig2}(j), with its central portion is illustrated in Figs.~\ref{fig:fig2}(k) and \ref{fig:fig2}(l). As depicted in Fig.~\ref{fig:fig2}(j), this Wigner function reveals more pronounced phase-space features: this particular arrangement of coherent states develops an octagon-like shape consisting of catlike interference fringes between the component states, along with a pronounced kaleidoscopic pattern that appears around the origin of the Poincar\'e\xspace disk, which is further highlighted in Fig.~\ref{fig:fig2}(k), with the feature of interest marked by a black dot existing at the center of the pattern. A deeper investigation of this feature is conducted to determine its boundary and confirm its shape by identifying the region where it diminishes, as shown in Fig.~\ref{fig:fig2}(l). This analysis reveals that the feature has a perfect circular shape, with a noticeable improvement compared to the previous cases of four- and six-component compass states. For higher values of $k$ (represented by the red and purple curves in Fig.~\ref{fig:fig2}(l)), the feature becomes more tightly confined within the circular region. These isotropic sub-Planck features persist for higher-component SU(1,1) compass states as long as the number of components is even. This behavior is further demonstrated by the ten-component ($N=10$) and sixteen-component ($N=16$) compass states, as discussed in the Appendix.

The isotropic sub-Planck features introduced in this section exhibit symmetric phase-space extensions that scale as $1/k$ in all directions. For $k \gg 1$, these features become much smaller than the phase space extension of a coherent state. In the limit $k \gg 1$, the SU(1,1) coherent states reduce to two-mode squeezed number states $\ket{\zeta_0,\Delta}$ (Eq.~(\ref{eq:coherent_state_twophotons})). This implies that the SU(1,1) superpositions introduced in this section (with $k\gg1$) are the superpositions of two-mode squeezed number states with their Bargmann index $k$ proportional to $\Delta$ (i.e., $k\propto \Delta$), where $\Delta$ represents the photon number asymmetry between the two correlated modes of the two-mode squeezed number states. Higher $k$ causes greater asymmetry in the photon number of two coupled modes of the two-mode squeezed number states. In other words, this increased asymmetry favors the formation of sub-Planck features in the corresponding SU(1,1) compass states.

\begin{figure*}[htp!]
\centering
\includegraphics[width=0.95\textwidth]{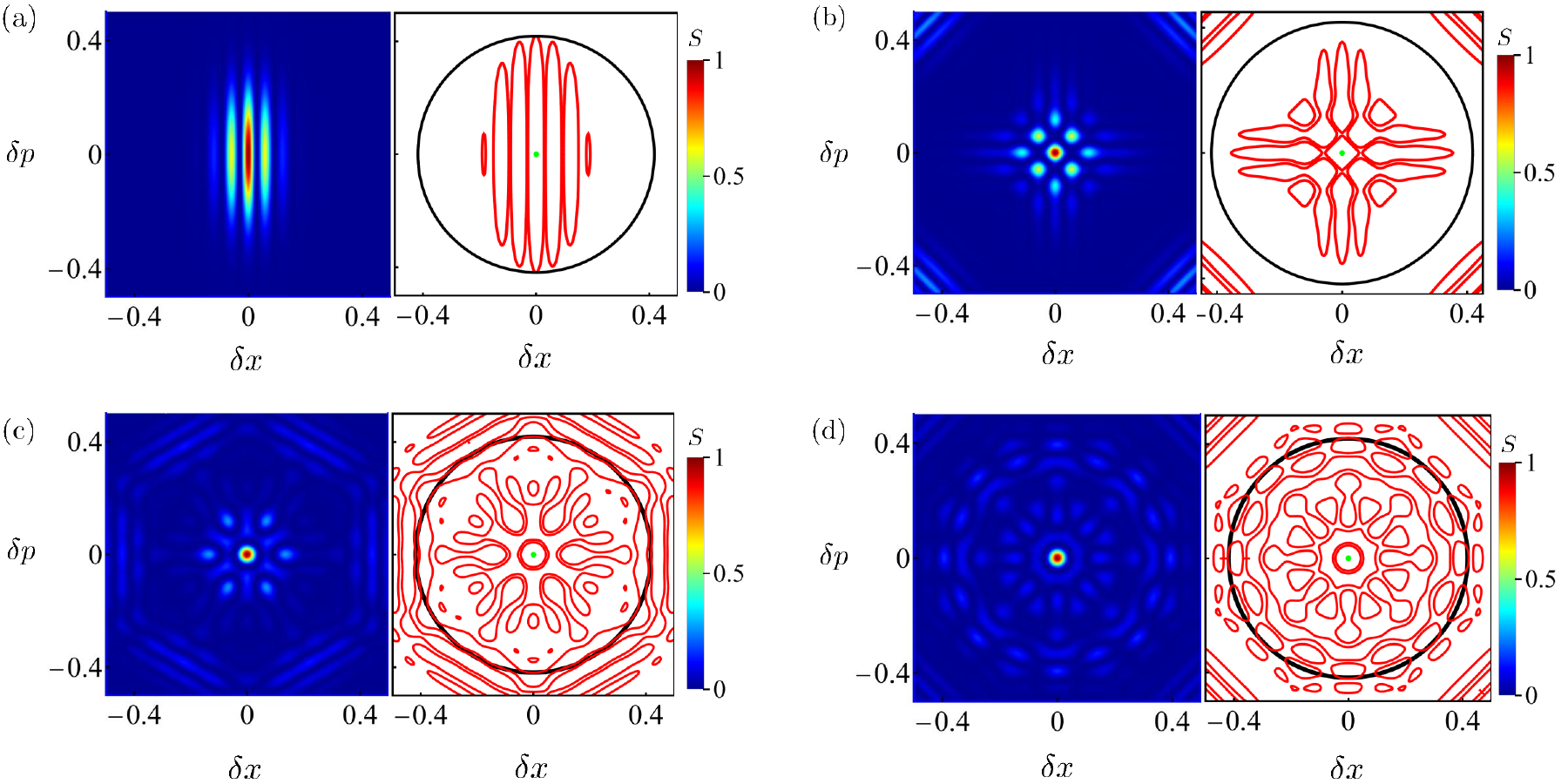}
\caption{The left panels show the overlap $S_{\ket{\bigcirc_{N}}}(\delta)$ with $\delta = \delta x + \text{i} \delta p$, where the corresponding zeros illustrated in the right panels. (a) the SU(1,1) cat state; (b) $N=4$; (c) $N=6$; (d) $N=8$. In all cases, $\overline{\tau}=1.5$ and $k$=12}.
\label{fig:fig3}
\end{figure*}
\subsection{Displacement sensitivity}

The enhancement in the sensitivity of a state to displacements, dedicated to its sub-Planck features, can be quantified by examining its orthogonality under infinitesimal shifts~\cite{Toscano06}; this approach is extended now for our SU(1,1) superpositions. The overlap between a state and its slightly displaced versions analyzes a state's response to displacements and eventually characterizes sensitivity, as given in Eq.~(\ref{eq:wig_sens}). Specifically, the overlap $S_{\ket{\bigcirc_{N}}}(\delta)= |\braket{\bigcirc_{N}|\hat{D}(\delta)|\bigcirc_{N}}|^2$, which directly associates the corresponding Wigner phase space, and its associated orthogonality $S_{\ket{\bigcirc_{N}}}(\delta) \simeq 0$ reflect the state’s response to displacement (sensitivity to phase-space displacements). In our case, this overlap reads

\begin{align}\label{eq:generic_ov}
  S_{\ket{\bigcirc_{N}}}(\delta)= \left| \sum_{i,j=0}^{N-1} S_{\ket{\zeta_i}\bra{\zeta_j}}(\delta)\right|^2
\end{align}
with
\begin{align}
  S_{\ket{\zeta_i}\bra{\zeta_j}}(\delta)= &\nonumber\text{exp}\left({-2\text{i}k\text{arg}(1+\delta^* \zeta_j)}\right) \\&\times\left[\frac{(1-|\zeta_i|^2)(1-\left|\frac{\delta+\zeta_j}{1+\delta^*\zeta_j}\right|^2)}{(1-\zeta^*_i)(\frac{\delta+\zeta_j}{1+\delta^*\zeta_j})}\right]^k.
\end{align}
Fig.~\ref{fig:fig3} illustrates the corresponding overlap $S_{\ket{\bigcirc_{N}}}(\delta)$ for the SU(1,1) quantum states of interest, extending the discussion from well-known states to newly introduced multicomponent superpositions. The overlap function between the SU(1,1) coherent state $\ket{0,k}$ and its displaced version $\hat{D}(\delta)\ket{0,k}$ is given by $S_{\ket{0,k}}(\delta) = \text{e}^{-k |\delta|^2}$, with its vanishing regions represented by a black circle, which appears alongside each illustration in Fig.~\ref{fig:fig3}. The extension of this circle is proportional to $\frac{1}{\sqrt{k}}$, which measures the amount of displacement required for the SU(1,1) coherent state to achieve orthogonality and predicts the sensitivity of SU(1,1) coherent states (shot-noise limit). This can be considered as the minimum norm (or SQL) against which the sensitivity of each state is compared.

We now examine the overlap function $S_{\ket{\bigcirc_{N}}}(\delta)$ for multiple $N$ values associating our SU(1,1) superpositions. Starting with the overlap function associated with the SU(1,1) cat state~\cite{naeem2022} [Fig.~\ref{fig:fig2}(a)], which is shown in Fig.~\ref{fig:fig3}(a) (left panel). We further illustrate its vanishing regions in Fig.~\ref{fig:fig3}(a) (right panel). The central feature in Fig.~\ref{fig:fig3}(a) (right panel), marked by the green dot, illustrates that along the $p$ direction in phase space, a displacement $\delta x \simeq \frac{1}{k}$ (which is $\frac{1}{\sqrt{k}}$ smaller than that of the coherent state) can make the SU(1,1) cat state orthogonal to its $\delta$-perturbed version. In contrast, along the $p$ direction, a displacement $\delta p$, of the same order as that of a coherent state (as observed in the central feature of the right panel of Fig.~\ref{fig:fig3}(a), where the central feature marked by the green dot coincides with the level of the coherent states) reaches distinguishability for a displacement proportional to $\frac{1}{\sqrt{k}}$. This demonstrates that the cat state is responsive to displacements smaller than the SQL along a specific direction, achieving an anisotropic enhancement in sensitivity. 

The case with $N = 4$ in Eq.~(\ref{eq:generic_ov}) determines the sensitivity of the SU(1,1) compass state, as shown in Fig.~\ref{fig:fig3}(b). When compared to the SU(1,1) cat state, the overlap of the compass state can attain orthogonality for displacements $\delta \simeq \frac{1}{k}$ nearly along any arbitrary direction in phase space. This suggests directional independence in sensitivity enhancement, providing advantages over cat states. However, as observed, the distinguishability occurs in tile-like regions (see the central region of Fig.~\ref{fig:fig3}(b) in the right panel, which is marked with a green dot), revealing the nonuniformity of displacements causing orthogonality of the state. This indicates the imperfections associated with the sensitivity of state sensitivity in detecting such minor displacements, as the state's sensitivity is not perfectly isotropic. Some directions offer more enhancements than others.

We observe that the anisotropic imperfections in the sensitivity of the SU(1,1) compass state can be progressively corrected by incorporating more components of coherent states in our superpositions, ultimately leading to perfect isotropic sensitivity enhancement and making them more desirable than their lower-order counterparts. For $N = 6$ in Eq.~(\ref{eq:generic_ov}), indicating the overlap between a six-component compass state and its displaced version, as depicted in Fig.~\ref{fig:fig3}(c), where the left panel represents the overlap function and its associated zero-amplitude regions are shown in the right panel, evidencing orthogonality at much smaller displacement compared to the coherent state. Notably, this distinguishability is achieved for displacement proportional to $\frac{1}{k}$, occurring in a more isotropic manner compared to the compass state, as evidenced by the circular region marked with a green dot in Fig.~\ref{fig:fig3}(c) (right panel). Our eight-component compass state achieved enhanced sensitivity to displacement compared to its lower-order counterparts. As shown in Fig.~\ref{fig:fig3}(d), the overlap and its associated orthogonality are concentrated in a more tightly confined isotropic region, represented by the green-dotted circular area. This indicates that a superposition of eight coherent states provides a greater isotropic improvement in displacement sensitivity compared to the lower-order superpositions examined in this study. Extending these superpositions to higher $N$ values preserves this enhanced sensitivity with even greater precision. The confinement of the associated overlap function to a circular path in phase space becomes evident for higher $N$, with the radius of the circular path proportional to $\frac{1}{k}$. The sensitivity to displacements associating ten- and sixteen-component SU(1,1) compass states is discussed in the Appendix.

\subsection{Analogues of SU(1,1) superposition}

We now construct analogous states of the superpositions defined in Eq.~(\ref{eq:su11_generic}) within a specific Hamiltonian framework. In particular, we begin with the Hamiltonian~\cite{quantum7030031}
\begin{align}
    \hat{H} = \eta_1 (\hat{a}_1^\dagger \hat{a}_1)^2 + \eta_2 (\hat{a}_2^\dagger \hat{a}_2)^2 + \eta_3 \, \hat{a}_1^\dagger \hat{a}_1 \, \hat{a}_2^\dagger \hat{a}_2,
    \label{eq:hami1}
\end{align}
which represents the Kerr-type interactions of two bosonic modes, 1 and 2, in the interaction picture. Using the SU(1,1) generators $\hat{K}_+$, $\hat{K}_-$, and $\hat{K}_0$, Hamiltonian (\ref{eq:hami1}) transforms to
\begin{equation}\label{eq:compact_hami}
    \hat{H} = \alpha_1 \hat{K}_0^2 + \alpha_2 \hat{\upsilon}^2,
\end{equation}
where $\hat{\upsilon} = \hat{a}_1^\dagger \hat{a}_1 - \hat{a}_2^\dagger \hat{a}_2, \quad [\hat{\upsilon}, \hat{K}_i] = 0$, and $i$ label different generators of the SU(1,1) group. Omitting the second term in the Hamiltonian~(\ref{eq:compact_hami}) corresponds to the compact case, under which the temporal evolution of the SU(1,1) coherent state $\ket{\zeta_0,k}$ can generate analogues of SU(1,1) compass states at particular instants of time.

\begin{figure}[htp!]
\centering
\includegraphics[width=0.451\textwidth]{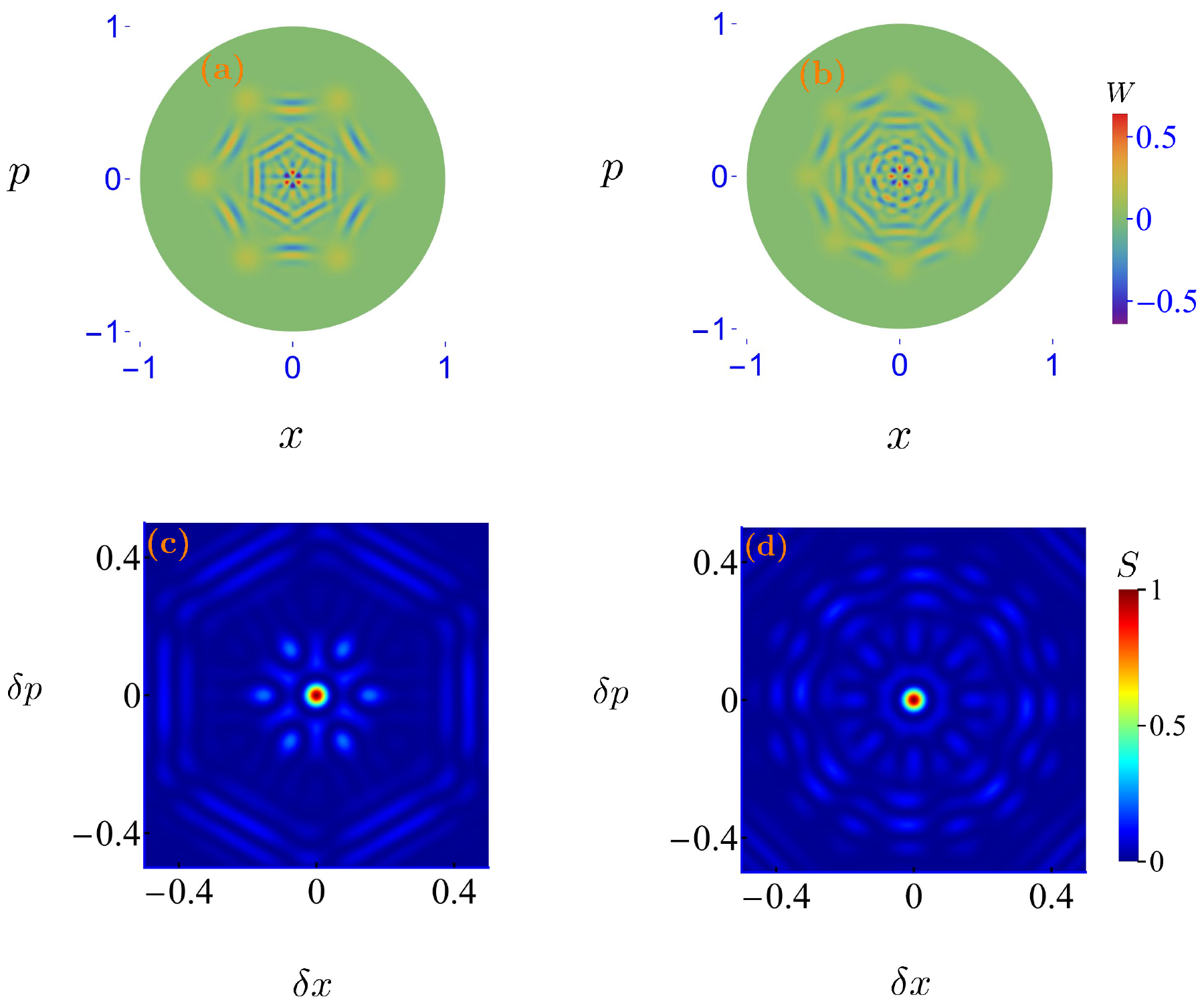}
\caption{Wigner distributions of the analogous six- and eight-component compass states: (a) $\ket{\psi_{6}}$ and (b) $\ket{\psi_{8}}$, with their corresponding overlaps shown in (c) and (d). In all cases, $\zeta_0=0.6$ and $k=12$.}
\label{fig:fig4}
\end{figure}

More explicitly, the evolved state under the compact form of the Hamiltonian~(\ref{eq:compact_hami}) is given by $\ket{\psi(t)}=\text{e}^{-\text{i}t\hat{K}_0^2}\ket{\zeta_0,k}$, which at $t=\nicefrac{\pi}{M}$ evolves into multicomponent SU(1,1) superpositions of the form $\ket{\psi_M}=\mathrm{e}^{- 2 \pi \mathrm{i} (\frac{1}{2 M}) \hat{K}_0^2} \ket{\zeta_0,k}$. This results in the following superposition state:
\begin{align}
\ket{\psi_M} =&\nonumber \left(1-|\zeta_0|^2\right)^k \sum_{n=0}^{\infty}\sqrt{\frac{\Gamma(n+2 k)}{n!\Gamma(2 k)}} \mathrm{e}^{- 2 \pi \mathrm{i} \frac{ \left( k+n\right)^2 }{2 M} }\\&\times  \zeta_0^n\ket{k, k+n},
\label{eq:generic_2}
\end{align}
where the quadratic phase factor in $n$ can be rewritten using the discrete Fourier transform (DFT)~\cite{naeemprapplied}. To understand it, let us consider a quadratic-phase function denoted as
\begin{equation}  
    g(n)= \exp\left(-2\pi\mathrm{i}\,\frac{p_1}{q_1}\,n^2\right),
\end{equation}  
with $p_1$ and $q_1$ being coprime integers. Let us assume that $g(n)$ is periodic with a fundamental period of $N$, i.e., the smallest positive integer satisfying $g(n+N) = g(n)$ for all integers $n$, and then employing DFT implies
\begin{equation}  
    g(n)= \exp\left(-2\pi\mathrm{i}\,\frac{p_1}{q_1}\,n^2\right) \overset{\text{DFT}}{=} \sum_{s=0}^{N-1}b_{s}\exp\left(-2\pi\mathrm{i}\,\frac{s}{N}\,n\right).
\end{equation}
This allows Eq.~(\ref{eq:generic_2}) to be expressed as multicomponent SU(1,1) superpositions of the form:
\begin{align}\label{eq:multi_gener}
    \ket{\psi_M}=\sum_{s=0}^{N-1} b_{s}\Ket{\zeta \mathrm{e}^{- 2 \pi \mathrm{i} \left(\frac{s}{N} +  \frac{k }{M} \right)}}.
\end{align}
where
\begin{equation}
b_s=\frac{1}{N} \sum_{n=0}^{N-1} g(n) \mathrm{e}^{+2 \pi \mathrm{i} \frac{n}{N} s}=\frac{1}{N} \sum_{n=0}^{N-1} \mathrm{e}^{2 \pi \mathrm{i}\left(\frac{n}{N} s-\frac{p_1}{q_1} n^2\right)}.
\end{equation}
Employing $p_1 = 1$ and $q_1 \equiv 2 \ (\mathrm{mod}\ 4)$, the coefficients ($b_s$) satisfy
\begin{equation}
b_s = \text{e}^{-2\pi \text{i} \frac{s}{N}} \, \text{e}^{-2\pi \text{i} \frac{1}{q_1}} \, b_{(s+2)\,(\mathrm{mod}\, N)},
\label{eq:iter1}
\end{equation}
while other choices of $q_1$ lead to
\begin{equation}
b_s = \text{e}^{-2\pi \text{i} \frac{s}{N}} \, \text{e}^{-2\pi \text{i} \frac{1}{q_1}} \, b_{(s+1)\,(\mathrm{mod}\, N)},
\label{eq:iter2}
\end{equation}
which determine the full set of coefficients $b_s$.

In Eq.~(\ref{eq:multi_gener}), particular choices of $M$ and $N$ yield the SU(1,1) superpositions of interest. For example, $M=N=2$ produces a superposition of two SU(1,1) coherent states, while $M=N=4$ generates a superposition of four SU(1,1) coherent states. Higher-order superpositions, such as those consisting of six or eight SU(1,1) coherent states, can be obtained by setting $M=N=6$ and $M=N=8$, respectively. The six-component SU(1,1) superposition is given by
\begin{widetext}
\begin{align}
\lvert \psi_6 \rangle
= \frac{1}{\sqrt{6}} \Big[ &
\lvert \text{e}^{-\text{i}\pi k/3}\zeta_0 \rangle
+ \text{e}^{\text{i}\pi/6}\lvert \text{e}^{-\text{i}\pi (k+1)/3}\zeta_0 \rangle
+ \text{e}^{2\text{i}\pi/3}\lvert \text{e}^{-\text{i}\pi (k+2)/3}\zeta_0 \rangle \nonumber \\
& - \text{i}\,\lvert \text{e}^{-\text{i}\pi (k+3)/3}\zeta_0 \rangle
+ \text{e}^{2\text{i}\pi/3}\lvert \text{e}^{-\text{i}\pi (k+4)/3}\zeta_0 \rangle
+ \text{e}^{\text{i}\pi/6}\lvert \text{e}^{-\text{i}\pi (k+5)/3}\zeta_0 \rangle
\Big].
\end{align}
\end{widetext}

\begin{widetext}
Likewise, the eight component SU(1,1) superpositions result in
    \begin{align}
\lvert \psi_8 \rangle
= \frac{1}{\sqrt{8}} \Big[ &
\lvert \text{e}^{-\text{i}\pi k/4}\zeta_0 \rangle
+ \text{e}^{\text{i}\pi/8}\lvert \text{e}^{-\text{i}\pi (k+1)/4}\zeta_0 \rangle
+ \text{i}\,\lvert \text{e}^{-\text{i}\pi (k+2)/4}\zeta_0 \rangle
+ \text{e}^{-7\text{i}\pi/8}\lvert \text{e}^{-\text{i}\pi (k+3)/4}\zeta_0 \rangle \nonumber \\
& + \lvert \text{e}^{-\text{i}\pi (k+4)/4}\zeta_0 \rangle
+ \text{e}^{-7\text{i}\pi/8}\lvert \text{e}^{-\text{i}\pi (k+5)/4}\zeta_0 \rangle
+ \text{i}\,\lvert \text{e}^{-\text{i}\pi (k+6)/4}\zeta_0 \rangle
+ \text{e}^{\text{i}\pi/8}\lvert \text{e}^{-\text{i}\pi (k+7)/4}\zeta_0 \rangle
\Big].
\end{align}
\end{widetext}
We illustrate the Wigner functions and overlap distributions of these analogous six- and eight-component SU(1,1) compass states in Fig.~\ref{fig:fig4}, demonstrating the preservation of the crucial phase-space characteristics as their counterparts of Sec.~\ref{subsec:isotropic_ideal}, particularly their isotropic sub-Planck-scale sensitivity to phase-space displacements. Furthermore, this construction can be extended to higher-component SU(1,1) superpositions, such as the ten-component state $
|\psi_{10}\rangle
=
\frac{1}{\sqrt{10}}
\sum_{s=0}^{9}
b_s
\left|
\text{e}^{-2\text{i}\left(\frac{k+s}{10}\right)\pi}\zeta_0
\right\rangle ,
$
and the twelve-component state
$
|\psi_{12}\rangle
=
\frac{1}{2\sqrt{3}}
\sum_{s=0}^{11}
b_s
\left|
\text{e}^{-2\text{i}\left(\frac{k+s}{12}\right)\pi}\zeta_0
\right\rangle ,
$
where $b_s$ can be obtained using the iterative relations given in Eqs.~(\ref{eq:iter1}) and (\ref{eq:iter2}). This provides a particular physical realization of the SU(1,1) circular superpositions introduced in our work.

\begin{figure}[H]
\centering
\includegraphics[width=0.5\textwidth]{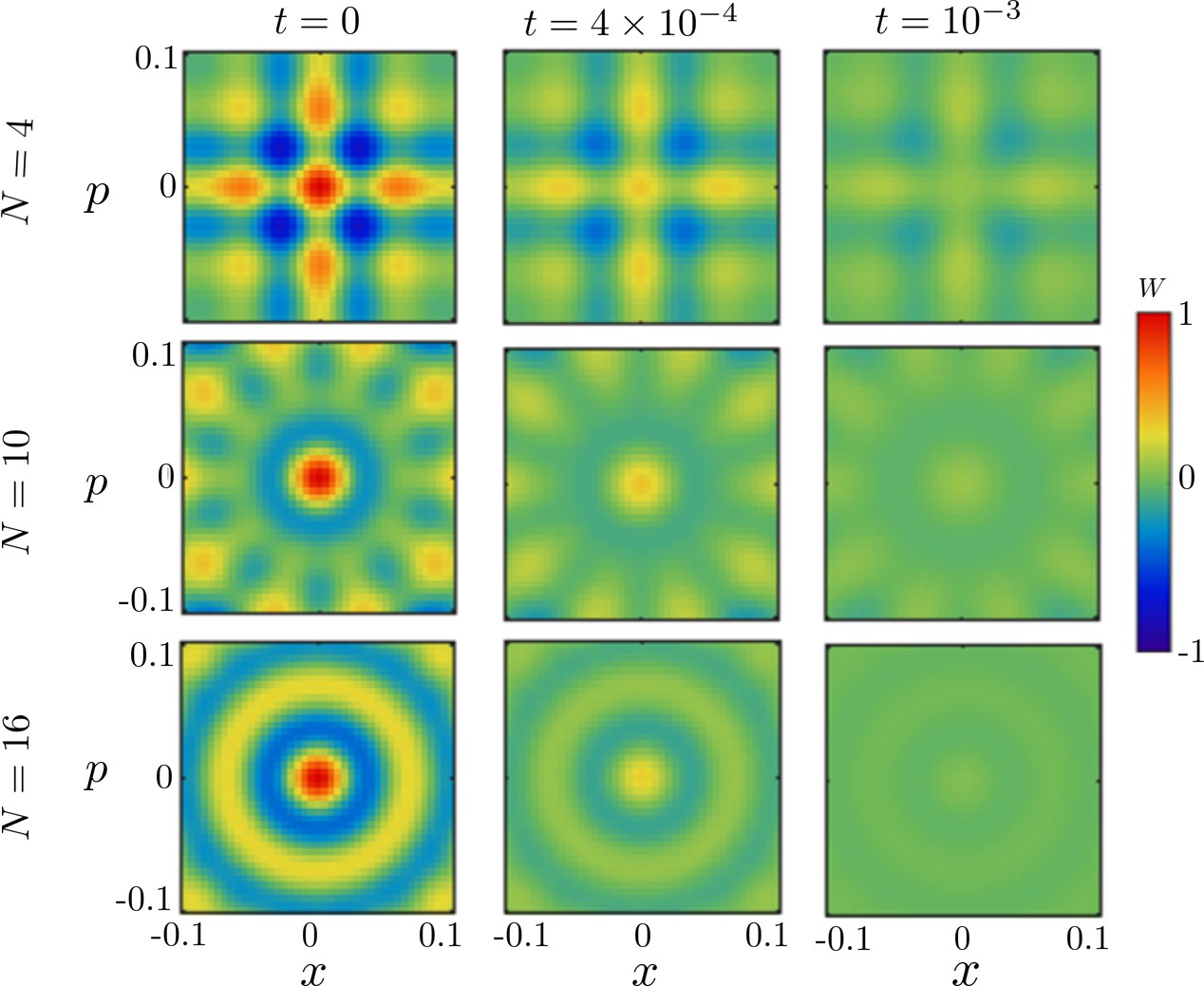}
\caption{Temporal evolution of the Wigner distributions of the corresponding evolved $N$-component SU(1,1) compass states under the impact of the frequency-resolved thermal Lindblad model [Eq.~(\ref{eq:decoh_model})]. As time ($t$) evolves, the associated nonclassical features gradually diminish under decoherence. In all cases, $k=12$ and $\overline{\tau}=1.5$.}
\label{fig:fig5}
\end{figure}

\begin{figure}[H]
\centering
\includegraphics[width=0.45\textwidth]{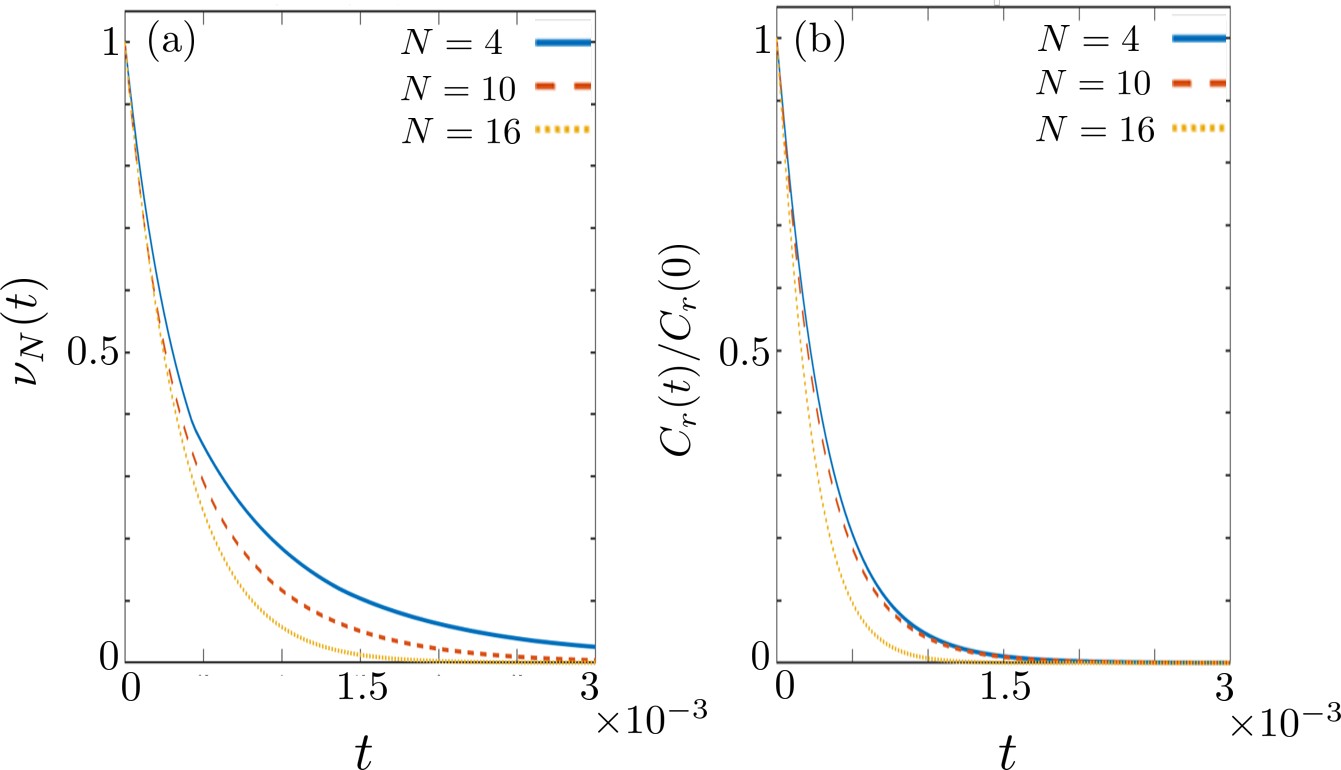}
\caption{(a) Central Wigner visibility $\nu_N(t)$ for SU(1,1) compass states with different numbers of components $N$; (b) normalized coherence $C_r(t)/C_r(0)$ for the same states. In all cases, $k=16$.}
\label{fig:fig6}
\end{figure}

\subsection{Stability under decoherence}\label{sec:outlook}
In this section, we examine the effect of decoherence on the crucial quantum features of SU(1,1) circular superpositions. To do this, we evolve these SU(1,1) superpositions $\hat{\rho}_N(0) = |O_N\rangle \langle O_N|$ (as defined in Eq.~(\ref{eq:su11_generic})), under a frequency-resolved thermal Lindblad model associated with the Hamiltonian $\hat{H}=\beta_1 \hat{K}_0$. Noting that $\hat{K}_0|k,n\rangle=(k+n)|k,n\rangle$, the corresponding energy spectrum is given by $E_n=\beta_1(k+n)$, which yields the transition frequencies $\Omega_n = E_{n+1} - E_n = \beta_1$. We use transition-dependent thermal occupations denoted as $\bar{N}_n = \frac{1}{\text{e}^{\beta_{\rm env} \Omega_n} - 1}$, where $\beta_{\rm env}$ is the inverse temperature of the environment.

We evolve the state according to
\begin{align}\label{eq:decoh_model}
\dot{\hat{\rho}}_N =&\nonumber -\text{i}[\hat{H},\hat{\rho}_N] + \sum_n \kappa (\bar{N}_n + 1) \mathcal{D}[A_n] \hat{\rho}_N \\&+ \sum_n \kappa \bar{N}_n \mathcal{D}[A_n^\dagger] \hat{\rho}_N,
\end{align}
where 
\begin{equation}
\mathcal{D}[L] \hat{\rho} = L \hat{\rho} L^\dagger - \frac{1}{2} \{ L^\dagger L, \hat{\rho} \},
\end{equation}
and $A_n = \sqrt{(n+1)(2k+n)} \, |k,n\rangle \langle k,n+1|$ is the lowering operator is associated with the transition $|k,n+1\rangle \to |k,n\rangle$, while $A_n^\dagger$ the reverse thermal excitation process is represented, and $\kappa=1$ throughout.
The Hamiltonian term produces coherent phase evolution of the off-diagonal density-matrix elements, while the dissipative ladder transitions suppress quantum coherence and redistribute the number-sector populations. This leads to a gradual decay of both the Wigner contrast and the relative entropy of coherence due to the thermal Lindblad dissipation, while the Hamiltonian governs the phase-space rotations and determines the transition frequency entering the bath occupation number. 

The loss of quantum features of the corresponding SU(1,1) superpositions is examined for a few representative cases, while the same trend persists for other associated superpositions. First, the gradual decay of quantum features over time is evident in Fig.~\ref{fig:fig5} for $N=4$, $N=10$, and $N=16$, highlighting the washout of central features. These superpositions provide coverage for both anisotropic sub-Planck features ($N=4$) and their anisotropic counterparts exhibited by the superpositions involving $N=10$ and $N=16$ components.

Further, the degradation of nonclassical features can be quantified through the variation between the maximum and minimum values of the corresponding Wigner distribution,
\begin{equation}
\Delta W_N(t)=
\max_{(x,p)\in R_o} W_{\ket{\bigcirc_N}}(x,p,t)
-
\min_{(x,p)\in R_o} W_{\ket{\bigcirc_N}}(x,p,t),
\end{equation}
where $R_o$ denotes the central region of phase space. The Wigner visibility function is defined as
\begin{equation}
\nu_N(t)=\frac{\Delta W_N(t)}{\Delta W_N(0)}.
\end{equation}
Similarly, the quantum coherence (the off-diagonal elements of the density matrix), which are responsible for developing interference fringes in phase space, can be quantified using the normalized relative entropy of coherence~\cite{PhysRevLett.113.140401}, that is,
\begin{equation}
\frac{C_r(t)}{C_r(0)} = \frac{C_r(\hat{\rho}_N(t))}{C_r(\hat{\rho}_N(0))}
\end{equation}
where
\begin{equation}
  C_r(\hat{\rho}) = S(\hat{\rho}_{d}) - S(\hat{\rho}),
\end{equation}
and $\hat{\rho}_{d}$ denote dephasing in the $|k,n\rangle$ basis. Both the Wigner visibility function and the relative entropy capture the decay of the corresponding interference terms in the Wigner function and clearly highlight the rate of this decay across different superpositions, as shown in Fig.~\ref{fig:fig6}. The results indicate that lower-order superpositions are comparatively more robust against decoherence, consistent with the behavior observed for analogous HW superpositions~\cite{naeemprapplied}. 
This suggests that isotropic sub-Planck structures, despite their enhanced potential for quantum metrology, may be more susceptible to decoherence compared to their anisotropic counterparts associated with lower-order compass states. Moreover, the decay of relative entropy aligns with the spectral picture provided by the entropic diagram and the entropy–SET representation, where decoherence manifests as increased mixedness and a higher effective spectral temperature~\cite{PhysRevResearch.7.023221,fhqs-g7c6}.

\section{SUMMARY}\label{sec:summary}

We have introduced superpositions involving $N$ components of SU(1,1) coherent states ($N$-component SU(1,1) compass states) to achieve an improved version of sub-Planck features. These resultant sub-Planck features are isotropic and are specifically achieved via the superpositions of $N \geq 6$ coherent states (with the total number limited to even integers). These component coherent states are uniformly distributed around the Poincar\'e\xspace disk. In other words, each coherent state is located at the same distance from the phase space origin and adjacent components are separated by a uniform angular spacing of $\frac{2 \pi}{N}$. More pronounced isotropic sub-Planck features emerge as the number of coherent states in the superpositions grows. These superpositions also exhibit isotropic enhancement in sensitivity to displacement, highlighting their increased significance for quantum metrology as compared to the standard SU(1,1) compass state~\cite{naeem2022}. We also presented the generation of these SU(1,1) superpositions within a Hamiltonian framework involving Kerr-type interactions between two bosonic modes, which also obey SU(1,1) dynamical symmetry. We have observed that the compact temporal evolution of the SU(1,1) coherent states under this Hamiltonian split into superpositions of $N$ SU(1,1) coherent states at a specific instant of time. Furthermore, we have analyzed the effect of decoherence on these SU(1,1) superpositions under a frequency-resolved thermal Lindblad model associated with a linear Hamiltonian. Our results highlight the limited stability of the corresponding SU(1,1) superpositions as the number of components $N$ increases, suggesting that isotropic sub-Planck features held heightened sensitivity to decoherence.

\subsection*{ACKNOWLEDGEMENT}

This work was supported by the National Natural Science
Foundation of China (Grants No.~12475009 and No.~12075001), the Anhui Province
Science and Technology Innovation Project (Grant No.~202423r06050004), the Anhui Provincial Department of Industry and Information Technology (Grant No.~JB24044), the Anhui Province Natural Science Foundation (Grant No.~2508085ZD001), and the Anhui Provincial University Scientific Research Major Project (Grant No.~2024AH040008). N.~A.~acknowledges the support of the National Natural Science Foundation of China (Grant No.~W2633009). J.-X.~P.~acknowledges support from the National Natural Science Foundation of China (Grant No. 12504566), the Basic Research Program of Jiangsu (Grant No.~BK20250947), the Natural Science Foundation of Jiangsu Higher Education Institutions (Grant No.~25KJB140013), and the Natural Science Foundation of Nantong City (Grant No.~JC2024045). T.~A.~acknowledges Mirpur University of Science and Technology (MUST), AJK, Pakistan, for granting ex-country leave.
\appendix

\section{Higher SU(1,1) superpositions and isotropy measure}\label{appendix:appendixA}
\begin{figure}[H]
\centering
\includegraphics[width=0.28\textwidth]{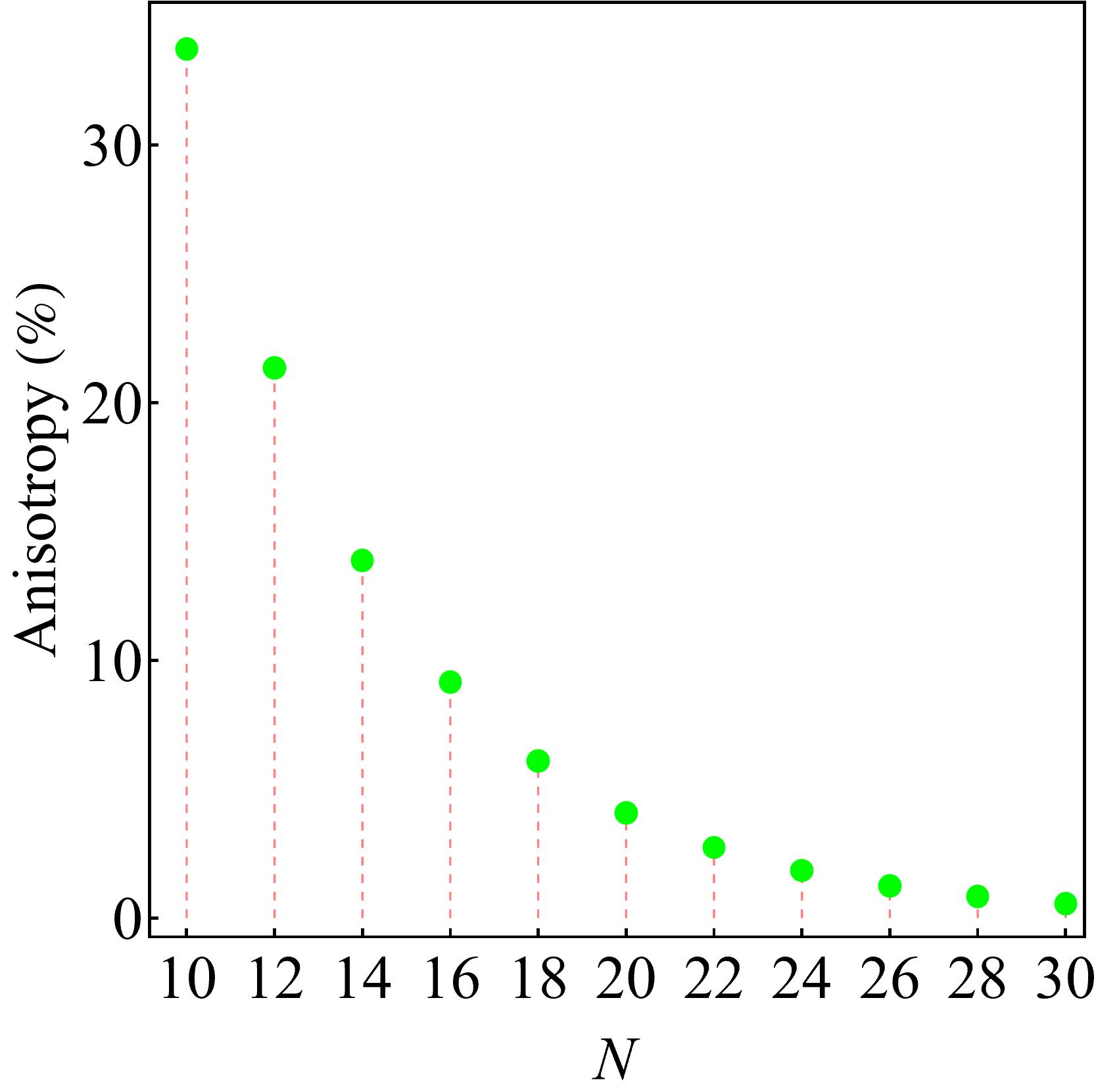}
\caption{Anisotropy of the central phase-space features of the Wigner distributions as a function of the number of superposition components ($N$) for SU(1,1) circular states. In all cases, $k = 12$ and $\overline{\tau} = 1.5$}.
\label{fig:fig7}
\end{figure}

\begin{figure}[H]
\centering
\includegraphics[width=0.41\textwidth]{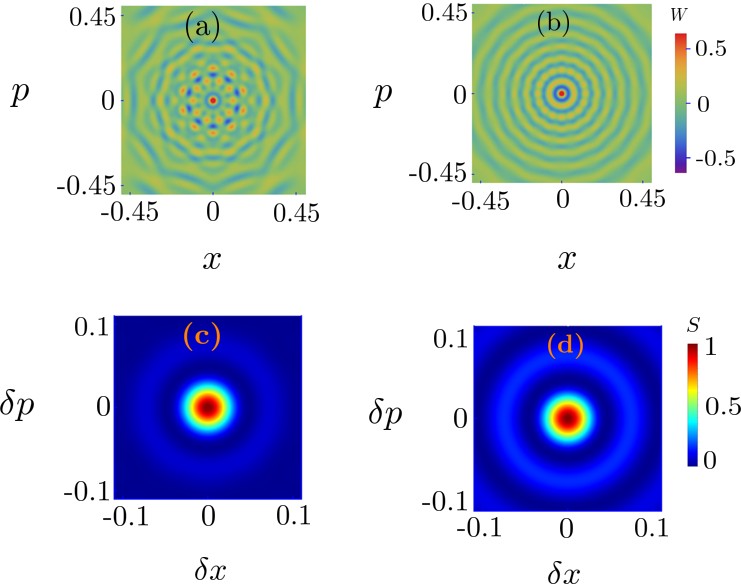}
\caption{The SU(1,1) Wigner distributions and their associated overlaps: (a) ten-component compass states and (b) sixteen-component compass states, with their corresponding overlaps shown in (c) and (d), respectively. In all cases, $k=12$ and $\overline{\tau}=1.5$.}
\label{fig:fig8}
\end{figure}

The superposition of SU(1,1) coherent states discussed in the main text includes the superposition up to $N=8$ components. More isotropically confined sub-Planck features are dedicated to the higher component superpositions, as discussed in the main text. This tendency is now confirmed for various higher-component SU(1,1) circular superpositions, providing more evidence of this claim. In this section, we evaluate the amount of anisotropy included in the core features by setting a particular threshold at which the related Wigner distribution meets this minor threshold, which extracts information about the percentage decrease in anisotropy as $N$ increases. 

To quantify the anisotropy with increased values of $N$, one may evaluate the boundary of each central patch by evaluating the Wigner functions reaching the threshold $\epsilon$, that is, $W_{|\bigcirc_N\rangle}(r(\phi)\cos\phi,r(\phi)\sin\phi) = \epsilon$ and $N\geq10$, and then assessing various radii along the different directions of the Poincar\'e\xspace disk, represented by $r(\phi)$. The anisotropy can be defined as the ratio of the maximum and minimum radii given by: Anisotropy$\sim\left(\frac{r_\text{max}}{r_\text{min}}-1\right)\times100$. This ratio is illustrated in Fig.~\ref{fig:fig7} for higher values of $N$, demonstrating that increasing the number of components in SU(1,1) circular states progressively reduces the anisotropy of their central phase-space features. In the limit of sufficiently large $N$, the central feature approaches a perfect circle, and the anisotropy consequently vanishes. This isotropic amplification can also be observed in their respective overlaps.

Here, we also illustrate the Wigner distribution and the overlaps for $N=10$ and $N=16$. As depicted, these higher-component superpositions generate more isotropic sub-Planck structures, which then result in greater isotropic enhancements in sensitivity to displacements, as shown in Fig.~\ref{fig:fig8}(a) ($N=10$) and Fig.~\ref{fig:fig8}(b) ($N=16$), with their corresponding overlaps shown in Figs.~\ref{fig:fig8}(c) and \ref{fig:fig7}(d), respectively. This shows that the isotropic characteristics in corresponding features of the states are expected to persist in higher-order superpositions, with further enhancement, provided that the angular spacing pattern is maintained and the number of components remains even.

\section*{DATA AVAILABILITY}
There are no publicly available research data or software supporting this manuscript. Requests for further
information or data should be sent to the authors.
.
\bibliography{ref}

@book{Per86,
title={Generalized Coherent States and Their Applications},
author={Perelomov, Askold},
series={Theoretical and Mathematical Physics},
publisher={Springer-Verlag},
address={Berlin},
year={1986}}

@article{ma2011quantum,
  title={Quantum spin squeezing},
  author={Ma, Jian and Wang, Xiaoguang and Sun, Chang-Pu and Nori, Franco},
  journal={Phys. Rep.},
  volume={509},
  number={2-3},
  pages={89--165},
  year={2011},
  doi={https://doi.org/10.1016/j.physrep.2011.08.003},
  publisher={Elsevier}
}

@article{Gerry01,
author = {Christopher C. Gerry},
journal = {Opt. Express},
number = {2},
pages = {76--85},
publisher = {Optica Publishing Group},
title = {Remarks on the use of group theory in quantum optics},
volume = {8},
month = {Jan},
year = {2001},
doi = {10.1364/OE.8.000076},
}

@article{Yazdi2008,
doi = {10.1088/1751-8113/41/5/055309},
year = {2008},
month = {jan},
publisher = {},
volume = {41},
number = {5},
pages = {055309},
author = {Shaterzadeh-Yazdi, Z and Turner, P S and Sanders, B C},
title = {{SU}(1,1) symmetry of multimode squeezed states},
journal = {J. Phys. A: Math. Theor.}
}

@article{Hudelist2014,
  title={Quantum metrology with parametric amplifier-based photon correlation interferometers},
  author={Hudelist, F and Kong, Jia and Liu, Cunjin and Jing, Jietai and Ou, ZY and Zhang, Weiping},
  journal={Nat. Commun.},
  volume={5},
  number={1},
  pages={1--6},
  year={2014},
  doi={10.1038/ncomms4049},
  publisher={Nature Publishing Group}
}

@article{Berrada2013,
  title = {Quantum metrology with {SU}(1,1) coherent states in the presence of nonlinear phase shifts},
  author = {Berrada, K.},
  journal = {Phys. Rev. A},
  volume = {88},
  issue = {1},
  pages = {013817},
  numpages = {6},
  year = {2013},
  month = {Jul},
  publisher = {American Physical Society},
  doi = {10.1103/PhysRevA.88.013817},
  url = {https://link.aps.org/doi/10.1103/PhysRevA.88.013817}
}

@article{Szigeti2017,
  title = {Pumped-Up {SU}(1,1) Interferometry},
  author = {Szigeti, Stuart S. and Lewis-Swan, Robert J. and Haine, Simon A.},
  journal = {Phys. Rev. Lett.},
  volume = {118},
  issue = {15},
  pages = {150401},
  numpages = {7},
  year = {2017},
  month = {Apr},
  publisher = {American Physical Society},
  doi = {10.1103/PhysRevLett.118.150401},
}

@article{barry2020,
  title = {Criticality in two-mode interferometers},
  author = {Liang, Hongbin and Su, Yuguo and Xiao, Xiao and Che, Yanming and Sanders, Barry C. and Wang, Xiaoguang},
  journal = {Phys. Rev. A},
  volume = {102},
  issue = {1},
  pages = {013722},
  numpages = {6},
  year = {2020},
  month = {Jul},
  publisher = {American Physical Society},
  doi = {10.1103/PhysRevA.102.013722},
}

@article{Gerry1991,
author = {Christopher C. Gerry},
journal = {J. Opt. Soc. Am. B},
number = {3},
pages = {685--690},
publisher = {OSA},
title = {Correlated two-mode {SU}(1,1) coherent states: nonclassical properties},
volume = {8},
month = {Mar},
year = {1991},
doi = {10.1364/JOSAB.8.000685},
}

@article{Seyfarth2020,
  doi = {10.22331/q-2020-09-07-317},
  url = {https://doi.org/10.22331/q-2020-09-07-317},
  title = {Wigner function for {SU}(1,1)},
  author = {Seyfarth, U. and Klimov, A. B. and Guise, H. de and Leuchs, G. and Sanchez-Soto, L. L.},
  journal = {{Quantum}},
  issn = {2521-327X},
  publisher = {{Verein zur F{\"{o}}rderung des Open Access Publizierens in den Quantenwissenschaften}},
  volume = {4},
  pages = {317},
  month = sep,
  year = {2020}
}

@article{Klimov2021,
	doi = {10.1088/1751-8121/abd7b4},
	url = {https://doi.org/10.1088/1751-8121/abd7b4},
	year = 2021,
	month = {jan},
	publisher = {{IOP} Publishing},
	volume = {54},
	number = {6},
	pages = {065301},
	author = {Andrei B Klimov and Ulrich Seyfarth and Hubert de Guise and Luis L S{\'{a}}nchez-Soto},
	title = {{SU}(1, 1) covariant s-parametrized maps},
	journal = {J. Phys. A: Math. Theor.}
}

@article{Gerry1995,
  title = {Two-mode intelligent {SU}(1,1) states},
  author = {Gerry, Christopher C. and Grobe, Rainer},
  journal = {Phys. Rev. A},
  volume = {51},
  issue = {5},
  pages = {4123--4131},
  numpages = {0},
  year = {1995},
  month = {May},
  publisher = {American Physical Society},
  doi = {10.1103/PhysRevA.51.4123},
}

@article{naeem2022,
  title = {Sub-{P}lanck phase-space structure and sensitivity for {SU}(1,1) compass states},
  author = {Akhtar, Naeem and Sanders, Barry C. and Xianlong, Gao},
  journal = {Phys. Rev. A},
  volume = {106},
  issue = {4},
  pages = {043704},
  numpages = {14},
  year = {2022},
  month = {Oct},
  publisher = {American Physical Society},
  doi = {10.1103/PhysRevA.106.043704},
  url = {https://link.aps.org/doi/10.1103/PhysRevA.106.043704}
}

@article{Linnemann_2017,
doi = {10.1088/2058-9565/aa802c},
url = {https://doi.org/10.1088/2058-9565/aa802c},
year = {2017},
month = {sep},
publisher = {IOP Publishing},
volume = {2},
number = {4},
pages = {044009},
author = {Linnemann, D and Schulz, J and Muessel, W and Kunkel, P and Prüfer, M and Frölian, A and Strobel, H and Oberthaler, M K},
title = {Active {SU}(1,1) atom interferometry},
journal = {Quantum Sci. Technol.}
}

@misc{morrison2023,
      title={Matrix {W}igner Function and {SU}(1,1)}, 
      author={P. G. Morrison},
      year={2023},
      eprint={2306.01238},
      archivePrefix={arXiv},
      primaryClass={quant-ph}
}

@article{h2x6-dz96,
  title = {Saturation of the quantum {C}ram\'er-{R}ao bound for distributed sensing via error sensitivity in {SU}(1,1)-{SU}($m$) interferometry},
  author = {Agarwal, Girish S.},
  journal = {Phys. Rev. A},
  volume = {112},
  issue = {3},
  pages = {032439},
  numpages = {9},
  year = {2025},
  month = {Sep},
  publisher = {American Physical Society},
  doi = {10.1103/h2x6-dz96}
}

@article{PhysRevA.51.1706,
  title = {Normal ordering of the {SU}(1,1) and {SU}(2) squeeze operators},
  author = {Lo, C. F.},
  journal = {Phys. Rev. A},
  volume = {51},
  issue = {2},
  pages = {1706--1708},
  numpages = {0},
  year = {1995},
  month = {Feb},
  publisher = {American Physical Society},
  doi = {10.1103/PhysRevA.51.1706},
  url = {https://link.aps.org/doi/10.1103/PhysRevA.51.1706}
}

@article{rv9m-n7px,
  title = {Realizing the optimal measurement scheme for an {SU}(1,1) interferometer in ab initio phase estimation},
  author = {Liu, Mingchen and Miao, Haixing and Zhang, Lijian},
  journal = {Phys. Rev. A},
  volume = {111},
  issue = {6},
  pages = {063711},
  numpages = {6},
  year = {2025},
  month = {Jun},
  publisher = {American Physical Society},
  doi = {10.1103/rv9m-n7px}
}

@article{naeemprapplied,
  title = {Multicomponent cat states with sub-{P}lanckian structures and their optomechanical analogues},
  author = {Hailin, Tan and Akhtar, Naeem and Xianlong, Gao},
  journal = {Phys. Rev. Appl.},
  volume = {24},
  issue = {2},
  pages = {024053},
  numpages = {20},
  year = {2025},
  month = {Aug},
  publisher = {American Physical Society},
  doi = {10.1103/gpj3-bbxq},
  url = {https://link.aps.org/doi/10.1103/gpj3-bbxq}
}

@article{Zurek2001,
title = {Sub-\uppercase{P}lanck structure in phase space and its relevance for quantum decoherence},
Author={Zurek, W. H. },
Journal = {Nature},
Pages = {712},
Volume =412,
Year = 2001,
doi = {10.1038/35089017}
}

@book{Sch01,
title =     {Quantum Optics in Phase Space},
author =    {W. P. Schleich},
publisher = {Wiley-VCH},
address={Weinheim},
year ={2001}}

@article{Howard2019,
  title = {Quantum hypercube states},
  author = {Howard, L. A. and Weinhold, T. J. and Shahandeh, F. and Combes, J. and Vanner, M. R. and White, A. G. and Ringbauer, M.},
  journal = {Phys. Rev. Lett.},
  volume = {123},
  issue = {2},
  pages = {020402},
  numpages = {5},
  year = {2019},
  month = {Jul},
  publisher = {American Physical Society},
  doi = {10.1103/PhysRevLett.123.020402},
  url = {https://link.aps.org/doi/10.1103/PhysRevLett.123.020402}}

@article{Toscano06,
  title = {Sub-\uppercase{P}lanck phase-space structures and \uppercase{h}eisenberg-limited measurements},
  author = {Toscano, F. and Dalvit, D. A. R. and Davidovich, L. and Zurek, W. H.},
  journal = {Phys. Rev. A},
  volume = {73},
  issue = {2},
  pages = {023803},
  numpages = {7},
  year = {2006},
  month = {Feb},
  publisher = {American Physical Society},
  doi = {10.1103/PhysRevA.73.023803},
  url = {https://link.aps.org/doi/10.1103/PhysRevA.73.023803}
}

@article{PhysRevA.69.052111,
  title = {Action scales for quantum decoherence and their relation to structures in phase space},
  author = {Alonso, Daniel and Brouard, S. and Palao, Jos\'e P. and Mayato, R. Sala},
  journal = {Phys. Rev. A},
  volume = {69},
  issue = {5},
  pages = {052111},
  numpages = {10},
  year = {2004},
  month = {May},
  publisher = {American Physical Society},
  doi = {10.1103/PhysRevA.69.052111},
  url = {https://link.aps.org/doi/10.1103/PhysRevA.69.052111}
}

@article{Robertson1929,
  title = {The Uncertainty Principle},
  author = {Robertson, H. P.},
  journal = {Phys. Rev.},
  volume = {34},
  issue = {1},
  pages = {163},
  numpages = {0},
  year = {1929},
  month = {Jul},
  publisher = {American Physical Society},
  doi = {10.1103/PhysRev.34.163},
  url = {https://link.aps.org/doi/10.1103/PhysRev.34.163}
}

@book{wheeler2014,
  title={Quantum Theory and Measurement},
  author={Wheeler, J. A.  and Zurek, W. H. },
  year={1983},
  publisher={Princeton University Press},
address={Princeton, NJ},
}

@article{Praxmeyer007,
  title = {Time-Frequency Domain Analogues of Phase Space Sub-{P}lanck Structures},
  author = {Praxmeyer, Ludmi\l{}a and Wasylczyk, Piotr and Radzewicz, Czes\l{}aw and W\'odkiewicz, Krzysztof},
  journal = {Phys. Rev. Lett.},
  volume = {98},
  issue = {6},
  pages = {063901},
  numpages = {4},
  year = {2007},
  month = {Feb},
  publisher = {American Physical Society},
  doi = {10.1103/PhysRevLett.98.063901},
  url = {https://link.aps.org/doi/10.1103/PhysRevLett.98.063901}
}

@article{Eff4,
title = {Quantum metrology at the \uppercase{h}eisenberg limit with ion trap motional compass states},
 author={Dalvit, D. A. R and de Matos Filho, R. L. and Toscano, F. },
Journal = {New J. Phys.},
pages = {276},
Volume =8,
Year = 2006,
doi ={10.1088/1367-2630/8/11/276}
}

@article{zheng2025,
  title = {Quantum-Enhanced Dark Matter Search Using Cat States},
  author = {Zheng, Pan and Cai, Yanyan and Xu, Bin and Wen, Shengcheng and Zhang, Libo and Ni, Zhongchu and Mai, Jiasheng and Zeng, Yanjie and Lin, Lin and Hu, Ling and Deng, Xiaowei and Liu, Song and Shu, Jing and Xu, Yuan and Yu, Dapeng},
  journal = {Phys. Rev. Lett.},
  volume = {136},
  issue = {17},
  pages = {171002},
  numpages = {8},
  year = {2026},
  month = {Apr},
  publisher = {American Physical Society},
  doi = {10.1103/wbhn-v1sw},
  url = {https://link.aps.org/doi/10.1103/wbhn-v1sw}
}

@article{deng2024quantum,
  title={Quantum-enhanced metrology with large {F}ock states},
  author = {X. Deng and S. Li and Z.-J. Chen and Z. Ni and Y. Cai and J. Mai and L. Zhang and P. Zheng and H. Yu and C.-L. Zou and et al.},
  journal={Nat. Phys.},
doi={https://doi.org/10.1038/s41567-024-02619-5},
volume={20},
  pages={1874–1880},
  year={2024},
  publisher={Nature Publishing Group UK London}
}

@article{PhysRevA.109.033724,
  title = {Generating overlap between compass states and squeezed, displaced, or {F}ock states},
  author = {Arman and Panigrahi, Prasanta K.},
  journal = {Phys. Rev. A},
  volume = {109},
  issue = {3},
  pages = {033724},
  numpages = {12},
  year = {2024},
  month = {Mar},
  publisher = {American Physical Society},
  doi = {10.1103/PhysRevA.109.033724}
}

@misc{arman2026enhancingsizephasespacestates,
      title={Enhancing the Size of Phase-Space States Containing Sub-{P}lanck-Scale Structures via Non-{G}aussian Operations}, 
      author={Arman and Prasanta K. Panigrahi},
      year={2026},
      eprint={2601.15654},
      archivePrefix={arXiv},
      primaryClass={quant-ph}
}

@article{llhr-hn2y,
  title = {Sub-{P}lanck structures and phase-space sensitivity of circular states},
  author = {Tang, Siting and Luo, Shunlong and Zhang, Yue},
  journal = {Phys. Rev. A},
  volume = {112},
  issue = {2},
  pages = {023712},
  numpages = {14},
  year = {2025},
  month = {Aug},
  publisher = {American Physical Society},
  doi = {10.1103/llhr-hn2y},
  url = {https://link.aps.org/doi/10.1103/llhr-hn2y}
}

@article{naeem2023,
  title = {Sub-{P}lanck structures and sensitivity of the superposed photon-added or photon-subtracted squeezed-vacuum states},
  author = {Akhtar, Naeem and Wu, Jizhou and Peng, Jia-Xin and Liu, Wu-Ming and Xianlong, Gao},
  journal = {Phys. Rev. A},
  volume = {107},
  issue = {5},
  pages = {052614},
  numpages = {18},
  year = {2023},
  month = {May},
  publisher = {American Physical Society},
  doi = {10.1103/PhysRevA.107.052614}
}

@misc{moore2026,
      title={Sub-{P}lanck structure quantification in non-{G}aussian probability densities}, 
      author={Darren W. Moore and Vojtěch Švarc and Kratveer Singh and Artem Kovalenko and Minh Tuan Pham and Ondřej Číp and Lukáš Slodička and Radim Filip},
      year={2026},
      eprint={2601.05898},
      archivePrefix={arXiv},
      primaryClass={quant-ph} 
}

@article{PhysRevLett.89.154103,
  title = {Decay of the {L}oschmidt Echo for Quantum States with Sub-{P}lanck-Scale Structures},
  author = {Jacquod, Ph. and Adagideli, I. and Beenakker, C. W. J.},
  journal = {Phys. Rev. Lett.},
  volume = {89},
  issue = {15},
  pages = {154103},
  numpages = {4},
  year = {2002},
  month = {Sep},
  publisher = {American Physical Society},
  doi = {10.1103/PhysRevLett.89.154103},
  url = {https://link.aps.org/doi/10.1103/PhysRevLett.89.154103}
}

@article{PhysRevA.93.053835,
  title = {Direct measurement of time-frequency analogs of sub-Planck structures},
  author = {Praxmeyer, Ludmila and Chen, Chih-Cheng and Yang, Popo and Yang, Shang-Da and Lee, Ray-Kuang},
  journal = {Phys. Rev. A},
  volume = {93},
  issue = {5},
  pages = {053835},
  numpages = {5},
  year = {2016},
  month = {May},
  publisher = {American Physical Society},
  doi = {10.1103/PhysRevA.93.053835},
  url = {https://link.aps.org/doi/10.1103/PhysRevA.93.053835}
}

@article{Naeem2021,
  title = {Sub-{P}lanck structures: Analogies between the {H}eisenberg-{W}eyl and {SU}(2) groups},
  author = {Akhtar, Naeem and Sanders, Barry C. and Navarrete-Benlloch, Carlos},
  journal = {Phys. Rev. A},
  volume = {103},
  issue = {5},
  pages = {053711},
  numpages = {14},
  year = {2021},
  month = {May},
  publisher = {American Physical Society},
  doi = {10.1103/PhysRevA.103.053711},
  url = {https://link.aps.org/doi/10.1103/PhysRevA.103.053711}
}

@article{akhtar2024sub-shot,
  title = {Sub-shot-noise sensitivity via superpositions of two deformed kitten states},
  author = {Akhtar, Naeem and Yang, Xiaosen and Peng, Jia-Xin and Ul Haq, Inaam and Xie, Yuee and Chen, Yuanping},
  journal = {Phys. Rev. A},
  volume = {111},
  issue = {3},
  pages = {032407},
  numpages = {17},
  year = {2025},
  month = {Mar},
  publisher = {American Physical Society},
  doi = {10.1103/PhysRevA.111.032407}
}

@article{Barry2023,
  title = {Superposing compass states for asymptotic isotropic sub-{P}lanck phase-space sensitivity},
  author = {Shukla, Atharva and Sanders, Barry C.},
  journal = {Phys. Rev. A},
  volume = {108},
  issue = {4},
  pages = {043719},
  numpages = {14},
  year = {2023},
  month = {Oct},
  publisher = {American Physical Society},
  doi = {10.1103/PhysRevA.108.043719}
}

@article{RevModPhys.90.035005,
  title = {Quantum metrology with nonclassical states of atomic ensembles},
  author = {Pezz\`e, Luca and Smerzi, Augusto and Oberthaler, Markus K. and Schmied, Roman and Treutlein, Philipp},
  journal = {Rev. Mod. Phys.},
  volume = {90},
  issue = {3},
  pages = {035005},
  numpages = {70},
  year = {2018},
  month = {Sep},
  publisher = {American Physical Society},
  doi = {10.1103/RevModPhys.90.035005},
  url = {https://link.aps.org/doi/10.1103/RevModPhys.90.035005}
}

@article{PhysRevLett.128.150501,
  title = {Metrological Characterization of Non-{G}aussian Entangled States of Superconducting Qubits},
  author = {Xu, Kai and Zhang, Yu-Ran and Sun, Zheng-Hang and Li, Hekang and Song, Pengtao and Xiang, Zhongcheng and Huang, Kaixuan and Li, Hao and Shi, Yun-Hao and Chen, Chi-Tong and Song, Xiaohui and Zheng, Dongning and Nori, Franco and Wang, H. and Fan, Heng},
  journal = {Phys. Rev. Lett.},
  volume = {128},
  issue = {15},
  pages = {150501},
  numpages = {8},
  year = {2022},
  month = {Apr},
  publisher = {American Physical Society},
  doi = {10.1103/PhysRevLett.128.150501},
  url = {https://link.aps.org/doi/10.1103/PhysRevLett.128.150501}
}

@article{7pyw-tgjd,
  title = {Optimal Phase-Insensitive Force Sensing with Non-{G}aussian States},
  author = {Grochowski, Piotr T. and Filip, Radim},
  journal = {Phys. Rev. Lett.},
  volume = {135},
  issue = {23},
  pages = {230802},
  numpages = {11},
  year = {2025},
  month = {Dec},
  publisher = {American Physical Society},
  doi = {10.1103/7pyw-tgjd},
  url = {https://link.aps.org/doi/10.1103/7pyw-tgjd}
}

@article{Audenaert14,
  author    = {K. M. R. Audenaert},
  title     = {Comparisons between quantum state distinguishability measures},
  journal   = {Quantum Inf. Comput.},
  volume    = {14},
  number    = {1-2},
  pages     = {31},
  year      = {2014},
  url       = {http://www.rintonpress.com/xxqic14/qic-14-12/0031-0038.pdf},
  bibsource = {dblp computer science bibliography, https://dblp.org}
}

@article{Martin_2020,
   title={Optimal Detection of Rotations about Unknown Axes by Coherent and Anticoherent States},
   volume={4},
   ISSN={2521-327X},
   DOI={10.22331/q-2020-06-22-285},
   journal={Quantum},
   publisher={Verein zur Forderung des Open Access Publizierens in den Quantenwissenschaften},
   author={Martin, John and Weigert, Stefan and Giraud, Olivier},
   year={2020},
   month=jun, pages={285} }

@article{Czartowski_2024,
   title={Minimal-noise estimation of noncommuting rotations of a spin},
   volume={8},
   ISSN={2521-327X},
   DOI={10.22331/q-2024-05-08-1341},
   journal={Quantum},
   publisher={Verein zur Forderung des Open Access Publizierens in den Quantenwissenschaften},
   author={Czartowski, Jakub and Życzkowski, Karol and Braun, Daniel},
   year={2024},
   month=may, pages={1341} }

@article{Sinatra2022,
    author = {Sinatra, Alice},
    title = {Spin-squeezed states for metrology},
    journal = {Appl. Phys. Lett.},
    volume = {120},
    number = {12},
    pages = {120501},
    year = {2022},
    month = {03},
    issn = {0003-6951},
    doi = {10.1063/5.0084096}
}

@article{Du2020,
    author = {Du, Wei and Chen, J. F. and Ou, Z. Y. and Zhang, Weiping},
    title = {Quantum dense metrology by an {SU}(2)-in-{SU}(1,1) nested interferometer},
    journal = {Appl. Phys. Lett.},
    volume = {117},
    number = {2},
    pages = {024003},
    year = {2020},
    month = {07},
    issn = {0003-6951},
    doi = {10.1063/5.0012304}
}

@article{PhysRevLett.113.140401,
  title = {Quantifying Coherence},
  author = {Baumgratz, T. and Cramer, M. and Plenio, M. B.},
  journal = {Phys. Rev. Lett.},
  volume = {113},
  issue = {14},
  pages = {140401},
  numpages = {5},
  year = {2014},
  month = {Sep},
  publisher = {American Physical Society},
  doi = {10.1103/PhysRevLett.113.140401},
  url = {https://link.aps.org/doi/10.1103/PhysRevLett.113.140401}
}

@article{PhysRevResearch.7.023221,
  title = {Entropic diagram characterization of quantum coherence: Degenerate distillation and the maximum eigenvalue uncertainty bound},
  author = {Aziz, Tariq and Song, Meng-Long and Ye, Liu and Wang, Dong},
  journal = {Phys. Rev. Res.},
  volume = {7},
  issue = {2},
  pages = {023221},
  numpages = {12},
  year = {2025},
  month = {Jun},
  publisher = {American Physical Society},
  doi = {10.1103/PhysRevResearch.7.023221},
  url = {https://link.aps.org/doi/10.1103/PhysRevResearch.7.023221}
}

@article{fhqs-g7c6,
  title = {Spectral bounds on entropy and ergotropy via statistical effective temperature in classical polarization and quantum thermal states},
  author = {Aziz, Tariq and Song, Meng-Long and Ye, Liu and Wang, Dong and Gil, Jos\'e J. and Kais, Sabre},
  journal = {Phys. Rev. Res.},
  volume = {7},
  issue = {3},
  pages = {033117},
  numpages = {18},
  year = {2025},
  month = {Aug},
  publisher = {American Physical Society},
  doi = {10.1103/fhqs-g7c6},
  url = {https://link.aps.org/doi/10.1103/fhqs-g7c6}
}

@article{PhysRevD.29.1107,
  title = {Impossibility of naively generalizing squeezed coherent states},
  author = {Fisher, Robert A. and Nieto, Michael Martin and Sandberg, Vernon D.},
  journal = {Phys. Rev. D},
  volume = {29},
  issue = {6},
  pages = {1107--1110},
  numpages = {0},
  year = {1984},
  month = {Mar},
  publisher = {American Physical Society},
  doi = {10.1103/PhysRevD.29.1107},
  url = {https://link.aps.org/doi/10.1103/PhysRevD.29.1107}
}

@article{Lawrie2029,
author = {Lawrie, B. J. and Lett, P. D. and Marino, A. M. and Pooser, R. C.},
title = {Quantum Sensing with Squeezed Light},
journal = {ACS Photonics},
volume = {6},
number = {6},
pages = {1307-1318},
year = {2019},
doi = {10.1021/acsphotonics.9b00250}
}

@article{Bu1990,
author = {Vladimír Bužek},
title = {{SU}(1,1) Squeezing of {SU}(1,1) Generalized Coherent States},
journal = {J. Mod. Opt.},
volume = {37},
number = {3},
pages = {303--316},
year = {1990},
publisher = {Taylor \& Francis},
doi = {10.1080/09500349014550371},


URL = { 
    
        https://doi.org/10.1080/09500349014550371
}
    
}

@inbook{SOLOMON,
author = {A. I. Solomon},
title = {Group Theory of superfluidity},
booktitle = {Dynamical Groups and Spectrum Generating Algebras},
Publisher = {World Scientific: Singapore, 1988},
pages = {523-527},
year={1988},
doi = {10.1142/9789814542319_0021},
URL = {https://www.worldscientific.com/doi/abs/10.1142/9789814542319_0021}
}

@Article{quantum7030031,
AUTHOR = {Aceves, Rodrigo D. and Baltazar, Miguel and Valtierra, Iván F. and Klimov, Andrei B.},
TITLE = {A Comparative Analysis of a Nonlinear Phase Space Evolution of {SU}(2) and {SU}(1,1) Coherent States},
JOURNAL = {Quantum Rep.},
VOLUME = {7},
YEAR = {2025},
NUMBER = {3},
DOI = {10.3390/quantum7030031}
}

@article{PhysRevLett.134.120201,
  title = {Phase-Space Measurements, Decoherence, and Classicality},
  author = {Brody, Dorje C. and Graefe, Eva-Maria and Melanathuru, Rishindra},
  journal = {Phys. Rev. Lett.},
  volume = {134},
  issue = {12},
  pages = {120201},
  numpages = {6},
  year = {2025},
  month = {Mar},
  publisher = {American Physical Society},
  doi = {10.1103/PhysRevLett.134.120201},
  url = {https://link.aps.org/doi/10.1103/PhysRevLett.134.120201}
}

@article{zurek_RevModPhys2003,
  title = {Decoherence, einselection, and the quantum origins of the classical},
  author = {Zurek, Wojciech Hubert},
  journal = {Rev. Mod. Phys.},
  volume = {75},
  issue = {3},
  pages = {715--775},
  numpages = {0},
  year = {2003},
  month = {May},
  publisher = {American Physical Society},
  doi = {10.1103/RevModPhys.75.715},
  url = {https://link.aps.org/doi/10.1103/RevModPhys.75.715}
}

@article{Akhtar2024,
  title = {Compasslike states in a thermal reservoir and fragility of their nonclassical features},
  author = {Akhtar, Naeem and Yang, Xiaosen and Asjad, Muhammad and Peng, Jia-Xin and Xianlong, Gao and Chen, Yuanping},
  journal = {Phys. Rev. A},
  volume = {109},
  issue = {5},
  pages = {053718},
  numpages = {13},
  year = {2024},
  month = {May},
  publisher = {American Physical Society},
  doi = {10.1103/PhysRevA.109.053718}
}

@article{akhtar2025D,
      title={Decoherence across phase-space scales: From compass states to general quantum states}, 
      author={Naeem Akhtar and Jia-Xin Peng and Tan Hailin and Xiaosen Yang and Dong Wang},
     journal = {Phys. Rev. A},
  volume = {113},
  issue = {6},
  pages = {062421},
  numpages = {},
  year = {2026},
  month = {Jun},
  publisher = {American Physical Society},
  doi = {10.1103/kgtv-6rmv},
  url = {https://doi.org/10.1103/kgtv-6rmv} 
}

@article{PhysRevLett.78.2547,
  title = {Method for Direct Measurement of the {W}igner Function in Cavity {QED} and Ion Traps},
  author = {Lutterbach, L. G. and Davidovich, L.},
  journal = {Phys. Rev. Lett.},
  volume = {78},
  issue = {13},
  pages = {2547--2550},
  numpages = {0},
  year = {1997},
  month = {Mar},
  publisher = {American Physical Society},
  doi = {10.1103/PhysRevLett.78.2547},
  url = {https://link.aps.org/doi/10.1103/PhysRevLett.78.2547}
}

@article{RevModPhys.81.299,
  title = {Continuous-variable optical quantum-state tomography},
  author = {Lvovsky, A. I. and Raymer, M. G.},
  journal = {Rev. Mod. Phys.},
  volume = {81},
  issue = {1},
  pages = {299--332},
  numpages = {0},
  year = {2009},
  month = {Mar},
  publisher = {American Physical Society},
  doi = {10.1103/RevModPhys.81.299},
  url = {https://link.aps.org/doi/10.1103/RevModPhys.81.299}
}

@article{Fadel_2025,
doi = {10.1088/1361-6633/ae00d8},
year = {2025},
month = {oct},
publisher = {IOP Publishing},
volume = {88},
number = {10},
pages = {106001},
author = {Fadel, Matteo and Roux, Noah and Gessner, Manuel},
title = {Quantum metrology with a continuous-variable system},
journal = {Rep. Prog. Phys.}
}

@article{ockeloen2018stabilized,
  title={Stabilized entanglement of massive mechanical oscillators},
  author={Ockeloen-Korppi, CF and Damsk{\"a}gg, E and Pirkkalainen, J-M and Asjad, M and Clerk, AA and Massel, F and Woolley, MJ and Sillanp{\"a}{\"a}, MA},
  journal={Nature},
  volume={556},
  number={7702},
  pages={478--482},
  year={2018},
  DOI={https://doi.org/10.1038/s41586-018-0038-x},
  publisher={Nature Publishing Group UK London}
}

@article{vlastakis2013deterministically,
  title={Deterministically encoding quantum information using 100-photon {S}chr{\"o}dinger cat states},
  author={Vlastakis, Brian and Kirchmair, Gerhard and Leghtas, Zaki and Nigg, Simon E and Frunzio, Luigi and Girvin, Steven M and Mirrahimi, Mazyar and Devoret, Michel H and Schoelkopf, Robert J},
  journal={Science},
  volume={342},
  number={6158},
  pages={607--610},
  year={2013},
  DOI={DOI: 10.1126/science.1243289},
  publisher={American Association for the Advancement of Science}
}

@article{leibfried2005creation,
  title={Creation of a six-atom ‘{S}chr{\"o}dinger cat’state},
  author={Leibfried, Dietrich and Knill, Emanuel and Seidelin, Signe and Britton, Joe and Blakestad, R Brad and Chiaverini, John and Hume, David B and Itano, Wayne M and Jost, John D and Langer, Christopher and others},
  journal={Nature},
  volume={438},
  number={7068},
  pages={639--642},
  year={2005},
  DOI={DOI
https://doi.org/10.1038/nature04251},
  publisher={Nature Publishing Group UK London}
}
 \end{document}